\documentclass{elsart}
\usepackage{latexsym}
\usepackage{epsfig}
\usepackage{natbib}

\def\url#1{#1} 

\begin{document}

\begin{frontmatter}

\title{Some implications of the introduction of scattered starlight 
in the spectrum of reddened stars.}

\author{Fr\'ed\'eric \snm Zagury}
\address{
\cty 02210 Saint R\'emy Blanzy, \cny France
\thanksref{email} }

   \thanks[email]{E-mail: fzagury@wanadoo.fr}

\received{July 2003}

 \begin{abstract}
This paper presents new investigations on coherent scattering
in the forward direction (orders of magnitude; conservation of energy; dependence of 
scattered light on geometry and wavelength), and on how 
scattered light contamination in the spectrum of reddened stars 
is possibly related to as yet unexplained observations (the diminution 
of the $2200\,\rm\AA$ bump when the obscuring material is close to the star, the 
difference between Hipparcos and photometric distances).
This paper then goes on to discuss the fit of the extinction curve, a possible 
role of extinction by the gas in the far-UV, and the reasons of 
the inadequacy of the \citet{fitz3} fit.
\end{abstract} 
\end{frontmatter}
 \section{Introduction} \label{intro}
When, in the past, observations were limited to the visible spectrum, 
interstellar extinction 
could be understood by the presence of large (compared to the 
wavelength) forward scattering grains, with scattering and extinction 
cross-sections proportional to $1/\lambda$; 
the visible extinction curve ($A_{\lambda}$ as a function of $1/\lambda$) 
is linear in $1/\lambda$.
The column density of large grains between a star and the observer 
is proportional to the slope of the visible extinction $A_{\lambda}$, 
which is twice the visible reddening $E(B-V)=A_{B}-A_{V}ее$ of the star if $\lambda$ is 
expressed in $\mu$m \citep{uv2}.

The problem of interstellar extinction was complicated by the 
release, in the 1960-70's, of the first UV spectra of reddened stars.
Extinction in the UV is less than the UV extension of the linear 
(in $1/\lambda$) rise of the visible extinction,
which can be explained either by a specific composition and size 
distribution of interstellar dust, or by an addition of 
starlight scattered and re-injected into the beam of the telescope.
In the former case we do see what we think we see, i.e. the direct light 
from the star extinguished by interstellar dust.
In the latter the effect of the additional component of scattered 
light diminishes the impression of extinction of the direct starlight by large grains.  

The UV extinction curve exhibits
a marked extinction feature, centered close to  $2200\,\rm\AA$,
in its near-UV part, and consists of a steep rise, though still under the 
UV extension of the visible extinction, in the far-UV.

If we observe the direct and extinguished light from reddened 
stars, the relation between these features and the large grains, and their 
variations from cloud to cloud, may be investigated by  
comparing extinction curves in different directions, 
normalized to a common slope in the visible
(i.e. to the same column density of large grains).
If extinction curves in different directions are normalized by $E(B-V)$, 
their visible parts superimpose well, but
large and unpredictible variations in the bump region and in the far-UV 
are observed (Figure~6 in \citet{bless72}):
for an equal column density of large grains, the observed UV 
extinction can noticeably differ from one interstellar cloud to another.

Consequently, if the light we receive from a reddened star is dominated 
by  the direct extinguished light from this star, large grains do 
not contribute to the UV extinction features, and their UV extinction 
must be a flat underlying continuum.
Different types of particles -a minimum of three- are then needed to explain 
the visible to UV extinction curve \citep{desert, greenberg}. 
Large grains are responsible for the extinction of starlight in the 
visible and have a grey extinction in the UV.
The $2200\,\rm\AA$ bump feature is attributed to a distribution of 
small grains, and the
far-UV rise of the extinction should be due to the presence in 
interstellar clouds of Polycyclic Aromatic 
Hydrocarbons (PAHs: \citet{allamandola85}) molecules, which should 
re-emitt in the near-infrared what they absorb in the far-UV.

The respective proportion of each of these three types of particles determines the 
importance of the extinction in each of the three parts 
(visible,  $2200\,\rm\AA$ bump, and 
far-UV regions) of the extinction curve.
The observed extinction curve in one direction is then the sum of these 
3 extinctions, and requires at least 3 parameters to be fitted.
In fact, \citet{fitz1,fitz2,fitz3} have shown that such a decomposition needs 
6 independent parameters.

However, this standard interpretation of the extinction curve meets 
observational and analytical difficulties.

The observed relationships between the different 
parts of the extinction curve  (see Figure~4 in \citet{savage85},
Figure~9 in \citet{savage75}, \citet{cardelli89}, and \citet{ uv6}), 
are difficult to reconcile with a three component grain model in 
which the proportion of each type of particles varies with the line of 
sight, and in an unpredictable way.
In addition, no dependence of the different types of particles 
on environment has ever been evidenced \citep{jenniskens93}, so that the 
reason for these variations remains a mystery.
For example, the claim that large grains might be destroyed into small 
grains in low column density, UV-exposed, regions, is 
contradicted by the linear relation which exists \citep{savage75} between the 
bump height and $E(B-V)$;
this relationship means that the column density of small grains 
must grow in proportion to that of large grains.
More specific problems remain: these include the precise identification of the PAH, 
the means by which they absorb the UV and re-emit in the near infrared,  
and the difficulty of reproducing in laboratory particles with such particular properties.

Recent analysis of UV observations by the International Ultraviolet 
Explorer (IUE) satellite 
contradict the near-UV breakdown of the extinction by large grains 
that is assumed by the standard theory.
The scattering cross-section deduced from the UV spectra of nebulae 
varies as $1/\lambda$ \citep{uv1}, as in the visible, which proves 
the continuity of the scattering cross section of interstellar 
matter, and a fortiori the continuity from the visible to the UV 
of the extinction law by large grains \citep{uv7}.

The visible-UV continuity of the extinction by large grains was further demonstrated 
for directions where the extinction is very low;
the light we receive from stars with $E(B-V)<<0.1$~mag. 
is extinguished by the same exponential 
of $1/\lambda$ in the visible and in the UV \citep{uv5}.
At intermediate reddenings ($\sim 0.1< E(B-V)<\sim 0.4$) 
the visible extinction law still extends in the near-UV, 
down to the bump region \citep{uv2}.

Analytically, the situation of the standard theory is no better.
Not only have \citet{fitz2} been unable to give any physical 
meaning to their mathematical expression of the far-UV extinction curve, 
but the \citet{fitz3} fit for extinction curves
is curiously confined to the UV domain: its
extension to the visible systematically diverges from the observations (see 
Figure~\ref{fig:fitz} of this paper and section~\ref{stfit} for a detailed 
discussion on this point).
Second (and it is stricking that this was passed over in silence 
in nearly all the articles published so far on the three component grain 
model)
\citet{cardelli89} have shown that extinction curves can generally be fitted 
using one parameter only, over all the visible-UV spectrum.
If one parameter is enough to reproduce a large set of extinction curves, why 
does the standard theory of extinction needs 6 parameters?
And why, with so many degrees of freedom, is it not able to do as 
well as the \citet{cardelli89} fit, i.e. to reproduce the complete 
visible+UV extinction curve?

Unless one has absolute faith (\citet{li}, `In Dust We 
Trust\ldots.') in the three component model, 
the several problems met by the standard theory of interstellar 
extinction prompt out to seek for other solutions.
The only alternative to the standard theory
is if the UV light we receive from reddened stars is not direct 
starlight alone, as it has been accepted so far, 
but that it also contains a non-negligible proportion of scattered light.
Scattered light will first modify the UV part of the spectrum because 
it is in the UV that extinction, thus the number of photons available 
for scattering, is the highest.

The scattered light component is most easily observed in directions 
of moderate reddening where it is limited to the far-UV wavelengths.
In these regions, it  clearly appears as an excess 
superimposed on the tail of the exponential extinction by the large grains \citep{uv2}.
At large optical depths, direct starlight, extinguished by large 
grains, becomes much smaller than scattered light.
This is especially true at UV wavelengths, where, if the linear 
extinction by large grains extends to the UV, a moderate column density 
of interstellar matter will be able to extinguish
all the direct starlight: we no longer see the star but its 
light scattered by foreground interstellar matter.
If the visible reddening $E(B-V)$ is important, the scattered light component can be 
large enough to modify the near-UV extinction, 
and even be detected in the visible \citep{uv3}.

The lower level of extinction in the UV, compared to what is expected 
by Mie theory for `normal' large grains, is due not to a change of the
extinction law of these grains between the visible and the UV, and to 
additional particles, but to this addition of scattered light.

This not only conforms to observations, but also leads to a simplification in our 
understanding of the interstellar medium:
large grains follow a continuous, linear, extinction law from the 
near infrared to the far-UV; their properties are the same in all directions.
The constraints on the nature of large grains implied by the 
standard theory, as well as the need for a grey extinction in the UV, are 
removed.

My papers published over the past five years have 
highlighted some properties of the scattered light component.
To explain both its importance, a probable 
dependence on $\lambda$ as $1/\lambda^{4}$, and the small angle within which it is 
effective, it seems that scattering must be coherent, thus 
confined to a very small angle (of order $10^{-8}$'') around the direction 
of the star, and due to spherically symetric particles \citep{uv8er}.
Hydrogen fullfills all the conditions imposed to the scatterers \citep{uv8er}.

In the first part of this paper I will reinvestigate some aspects of 
coherent scattering from the near forward directions (section~\ref{us} 
to section~\ref{geom}).
Amongst the consequences, I shall then discuss the extinction curve of stars with 
circumstellar dust (section~\ref{geo} to section~\ref{sitko}), 
the spurious effect a scattered light component could have
on distance estimates established by means of photometry (section~\ref{dis}), and the 
fit of the extinction curve (section~\ref{fit}).
\section{Incoherent scattering} \label{us}
\begin{figure}[]
\resizebox{0.5\columnwidth}{!}{\includegraphics{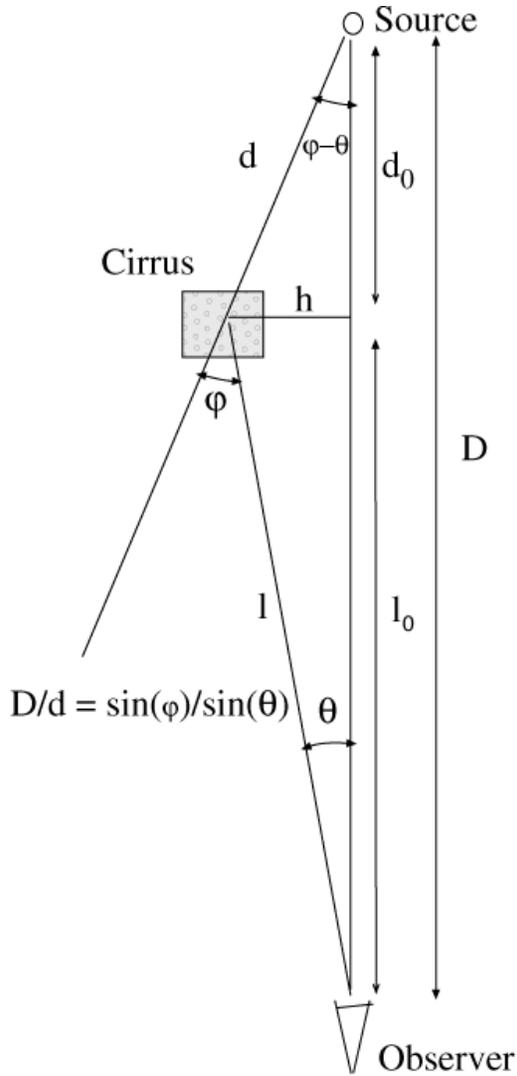}} 
\caption{Representation of the angles employed in the text} 
\label{fig:schema}
\end{figure}
Let us consider a star of luminosity $L_{\star}$ observed behind an interstellar 
cloud.
The cloud is at a mean distance $d_{0}$ from the star and $l_{0}$ from the 
observer. 
$D=d_{0}+l_{0}$ is the distance star-observer.
$d$ and $l$ are the distance of a point $M$ of the cloud, its 
distance to the star-observer axis is $h$ 
(see Figure~\ref{fig:schema}).

The star is assimilated to a point source of
unreddened flux measured by an observer on earth: $F_{0}=L_{\star}/(4\pi D^{2})$.

The power per unit surface received by the observer due to the 
scattering of starlight by one atom, with scattering cross section 
$\sigma$, at $M$ is:
\begin{equation}
  P_{u}е=\frac{\sigma L_{\star}}{\left(4\pi d^{2}\right)
  \left(4\pi l^{2}\right)}=
  \frac{\sigma D^{2}е}{4\pi d^{2}l^{2}} F_{0}
    \label{eq:pu}
\end{equation}е
[for sake of simplicity, since we will be interested by orders of 
magnitude only, the Rayleigh cross-section is assumed to be 
isotropic; this is true within a factor of two, see 
\citet{vandehulst}, section~6.12].

For an average column density of hydrogen $N_{H}$ of the cloud, 
the power scattered by the gaz reaching earth per unit surface can be 
estimated by: 
\begin{equation}
  P_{us}е=N_{H}\int_{0}^{\infty} е\frac{\sigma L_{\star}}
  {\left(4\pi d^{2}\right)
  \left(4\pi l^{2}\right)}2\pi h \mathrm{d} h,
    \label{eq:pun}
\end{equation}е
if the scattering is not coherent.

Considerations on energy conservation show that the power scattered by 
an homogenous medium and received per unit surface will not exceed $F_{0}$.
This is demonstrated by imagining a spherical sheet of gas centered on the 
star and dense enough to extinguish all of its light.
The scattered light received by a unit surface of a second sphere of 
radius $D$ will then be at its maximum and, for reasons of symetry, 
$P_{us}=L_{\star}/(4\pi D^{2})=F_{0}$.

$P_{us}$, for a cloud half way between the star and the 
observer ($l_{0}=d_{0}=D/2$), will be: 
\begin{eqnarray}
  P_{us}&=&\int_{0}^{\infty}
  N_{H} \frac{\sigma L_{\star}}
  {8\pi\left(l_{0}\cos \theta^{-1}е\right)^{4}}
  l_{0}^{2}еtg \theta \mathrm{d} (tg \theta) \nonumber \\
   &=& \sigma N_{H}е\frac{L_{\star}}{64\pi^{2} l_{0}^{2}}
   \int_{0}^{\infty}\frac{1}{\left( 1+x \right)^{2}}
  l_{0}^{2} \mathrm{d} x \nonumber       \\
&=& 0.5\sigma N_{H} F_{0}е\\
&=& 0.5\frac{\sigma_{0} N_{H}F_{0}е}{\lambda^{4}} \,
\label{eq:pun1}
\end{eqnarray}е
with $\sigma_{0}=\sigma \lambda^{4}$, and $x=tg^{2}\theta$е.

The Rayleigh cross-section of a particle small compared to the 
wavelength is $(8/3)\pi k^{4}\alpha^{2}$ \citep{vandehulst}, with 
$k=2\pi/\lambda$, and $\alpha$ the polarisability of the particle.
For atomic hydrogen, $\alpha=(9/2)a_{0}^{3}=6.7\,10^{-25}\,\rm cm^{3}$ 
(CGS units, $a_{0}=0.52\,\rm\AA$ is the Bohr radius 
of hydrogen).
The Rayleigh cross-section of hydrogen is:
$\sigma=5.86\,10^{-45}/\lambda^{4}\,\rm cm^{2}$ ($\lambda$ in cm).

Typical reddenings $E(B-V)$ of a few $0.1\,$~mag. ($A_{V}\sim$ a few 
0.3~mag.) correspond to column 
densities of hydrogen $N_{H}\sim 10^{21}\,\rm cm^{-2}е$ \citep{bohlin78}.
In the UV, $1/\lambda^{4}$ is in the range $10^{19}-10^{20}е\,\rm cm^{-4}$.
The maximum contribution of incoherent 
scattering by the gaz to the UV light received from the direction of the star, 
estimated from Equation~\ref{eq:pun1}, will remain small, $3\,10^{-4}F_{0}$ at most.
\section{Coherent scattering in the near forward directions} \label{cs}
It is only in the near forward directions that scattering is 
singular enough to require a specific attention since 
the waves scattered by all atoms on the 
line of sight, regardless of their spatial organisation, keep a 
phase relation \citep{bohren}.

This coherent scattering is superimposed on the incoherent scattering 
estimated above, and restricted to a very small volume of the cloud around the 
direction of the star.
As a first approximation, it will be assumed that the 
coherent scattered light can be described by a mean surface $S_{0}$ of the cloud around the 
direction of the star, and by $n_{c}$ atoms of the cloud within $S_{0}$, so that
the intensity of the coherent scattering the observer measures
is simply $n_{c}^{2}P_{u}$.

This assumption separates a focused beam of 
scattered light (received from the near-forward directions), from the 
classical incoherent scattered light (received from the rest of the cloud).
Reality is certainly more complex, with successive regions in phase and in 
phase opposition, as we move away from the direction of the star.
In a perfectly homogenous medium the contributions of 
these successive zones must cancel to leave only the incoherent 
component of scattered light.
The simplified approach adopted here will nevertheless be used to derive first orders of magnitude 
for $S_{0}$, $n_{c}$, and an upper value of the power $P_{cs}$ of the coherent 
scattered light received by the observer.

The order of magnitude of $S_{0}$ should be given by the size of the 
first Fresnel zones, for the wavelength range over which coherent 
scattering is observed.
The surface of the first Fresnel zone, as it is viewed by the observer, 
for a given wavelength $\lambda$, is: 
\begin{equation}
  S_{\lambda}е=\pi \lambda\frac{d_{0}l_{0}}{D},
    \label{eq:sl}
\end{equation}
$S_{\lambda}$ is maximum when $d_{0}=l_{0}=D/2$.

$S_{\lambda}$ defines the region of the cloud within which all the 
path-lengths from the star to the observer differ by less than $\lambda/2$.
It is also the largest surface within which the scattered light, at 
wavelength $\lambda$, 
observed at the positon of the observer, is constructively coherent.
Consequently, if $\lambda_{i}$ is a lower limit of the wavelength 
range over which scattering is constructively coherent: $S_{0}\leq S_{\lambda_{i}е}$.

For an atom in $S_{0}$, $\Delta$, the difference between the star-atom-observer 
path-length and $D$, is (see Figure~\ref{fig:schema}):
\begin{equation}
  \Delta=\frac{1}{2}\frac{h^{2}D}{l_{0}d_{0}}=\frac{1}{2}\theta^{2}\frac{Dl_{0}}{d_{0}}
    \label{eq:tradif}
\end{equation}е
Viewed from the position of the observer, $S_{0}$ is seen under an 
angle $\theta_{0}=(r_{0}/l_{0})^{0.5}е$ ($r_{0}$ is the radius of 
$S_{0}$), which must be less еthan the angle $\theta_{\lambda_{i}е}$:
\begin{equation}
    \theta_{\lambda_{i}} = \left(\frac{\lambda_{i}е d_{0}}{Dl_{0}}\right)^{0.5}=
    3\,10^{-8}"\left(\frac{\lambda_{i}е}{2000\,\mathrm{\AA}}\right)^{0.5}
     \left(\frac{100\,pc}{l_{0}}\right)^{0.5}
     \left(\frac{d_{0}}{D}\right)^{0.5}
    \label{eq:theta0}
\end{equation}
($\theta_{\lambda_{i}}$ was obtained by setting $\Delta=\lambda_{i}/2$ 
in Equation~\ref{eq:tradif}).е

$n_{c}$ can logically be estimated by $N_{H}еS_{0}$.

The power scattered by $S_{0}$, еat wavelength $\lambda$, is:
\begin{equation}
  P_{cs}=n_{c}^{2}P_{u}=
   \frac{N_{H}^{2}S_{0}^{2}D^{2}\sigma_{0}}
   {4\pi d_{0}^{2}l_{0}^{2}\lambda^{4}} F_{0}
     \label{eq:pc}
\end{equation}
An upper value of $P_{cs}$ is determined by setting 
$S_{0}=S_{\lambda}$:
\begin{equation}
  P_{cs,max}=
  \frac{\pi N_{H}^{2}\sigma_{0}}{4\lambda^{2}} F_{0}
     \label{eq:pcm}
\end{equation}
Note that $P_{cs}$ depends on $\lambda^{-4}$, while $P_{cs,max}$ 
varies as $\lambda^{-2}$.е

The power $P_{S_{0}е}$ extinguished by the gas in 
$S_{0}$ is:е
\begin{equation}
  P_{S_{0}е}=\sigma N_{H}е\frac{L_{\star}}{4\pi d_{0}^{2}}S_{0}
  =\sigma N_{H}S_{0}\frac{Dе}{d_{0}}^{2}F_{0}е
    \label{eq:ps0}
\end{equation}
The ratio, $P_{S_{0}е}/(r_{d}^{2}ееP_{cs})$, of the power extinguished 
in $S_{0}$ to the power received by a detector of size $r_{d}$ at the 
position of the observer is:ее
\begin{equation}
  \frac{P_{S_{0}е}}{r_{d}^{2} P_{cs}}е=
  \frac{4\pi l_{0}^{2}ее}{N_{H}S_{0}r_{d}^{2}е}
    \label{eq:prat}
\end{equation}
\section{Numerical orders of magnitude} \label{mag}
\subsection{Numerical application} \label{an}
To estimate orders of magnitude for the quantities defined in 
section~\ref{cs}, we will suppose $d_{0}=l_{0}=D/2$, and
a cloud, of average column density 
$N_{H}=10^{21}\rm cm^{-2}$ [$E(B-V)\sim 0.3$~mag.], 
a hundred parsec far from the sun.
The rayleigh cross-section of hydrogen at $\lambda=1500\,\rm\AA$ 
($1/\lambda=6\,\mu\rm m^{-1}$), the UV wavelength used in this 
section, is $\sigma=10^{-25}\,\rm cm^2$.
The low column density approximation applies since $\sigma N_{H}\sim 
10^{-4}<<1$.

The following orders of magnitude are then obtained: 
\begin{eqnarray}
    \theta _{\lambda} &=& 
    \left(\frac{\lambda}{2l_{0}}\right)^{0.5}
 \sim 3\,10^{-8}" \\
S_{\lambda}&=&\pi\lambda \frac{l_{0}}{2}\sim 7\,10^{15}\,\mathrm{cm}^{2}е\\
r_{\lambda}&=&\left(\pi ^{-1}S_{\lambda}\right)^{0.5}
=\left(\lambda \frac{l_{0}}{2}\right)^{0.5}\sim 5\,10^{7}\,\mathrm{cm}\\
n_{c}& \sim  & N_{H}S_{\lambda}е \sim 5\,10^{36} \mathrm{atoms}  \\
P_{cs,max}&=& \frac{\pi}{4} N_{H}^{2}е\lambda^{2}е\sigma F_{0}  \sim  
2\,10^{7}F_{0}
  \label{eq:mag}
\end{eqnarray}е
The radius of the first Fresnel zone, for astronomical distances, is extremely 
large, of order $500$~km, at UV wavelengths and for a cloud 100~pc away.
The power per unit surface of the light scattered from the first 
Fresnel zone, at the position of the observer, can theoretically be much larger 
than the unreddened flux of the star.
\subsection{Comparison with observations} \label{obs}
\begin{figure}[]
\resizebox{\columnwidth}{!}{\includegraphics{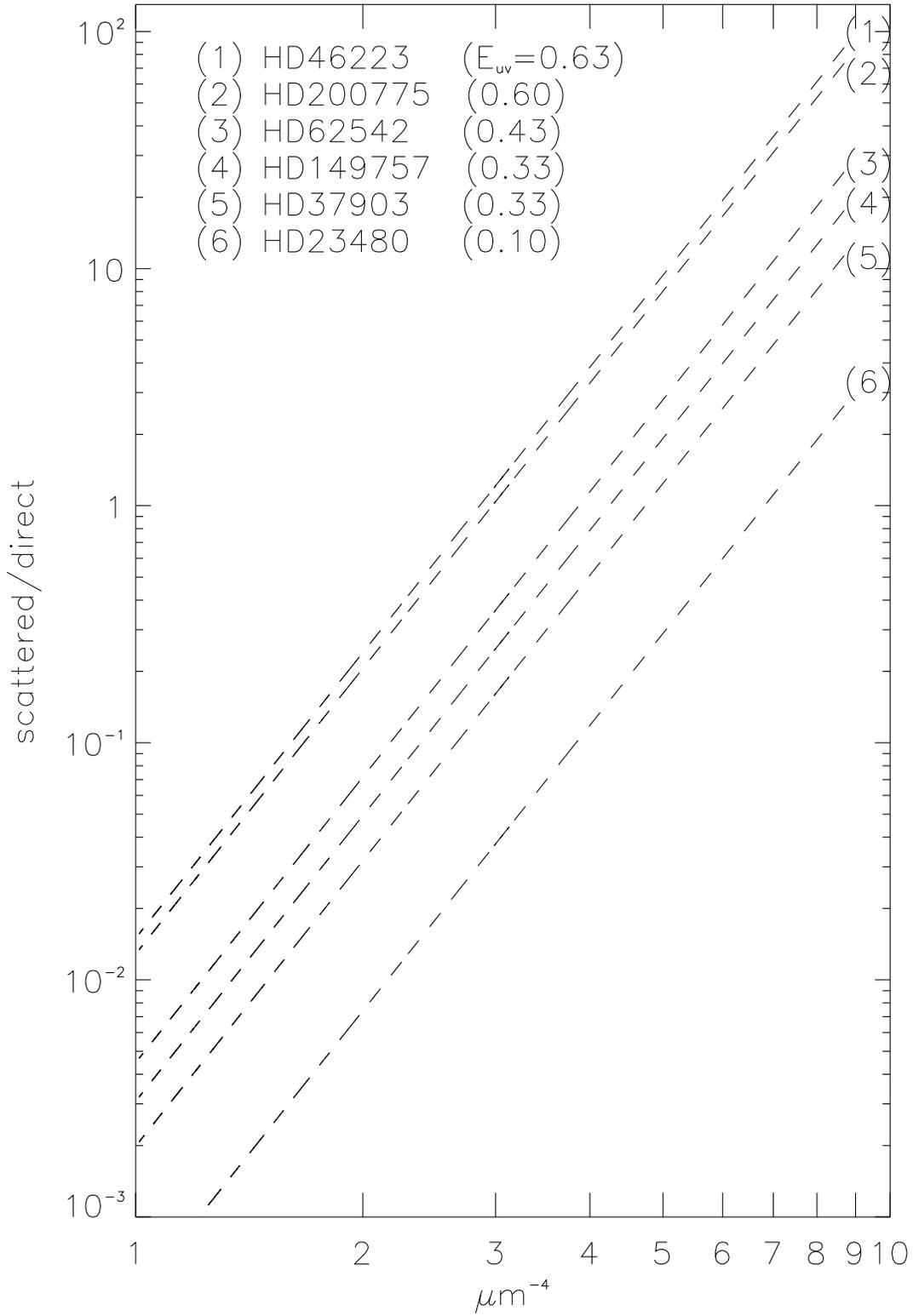}} 
\caption{Ratio, $P_{cs}/F_{0}$, of scattered to direct starlight for a sample of stars.
The ratios correspond to the $\beta E_{uv}^{2}/\lambda^{4}$ term of 
Equation~\ref{eq:fit1}.
They are deduced from Table~\ref{tbl:fit}, from \citet{uv3} 
for HD46223.
} 
\label{fig:rap}
\end{figure}
If extinction by large grains affects  scattered light in the same 
proportion  as direct starlight, and if 
the Rayleigh (exponential in $1/\lambda^{4}$) extinction by the gas remains small,
$P_{cs}/F_{0}е$ is equal to the observed ratio of scattered to direct starlights.

Figure~\ref{fig:rap} plots the ratio of scattered to direct 
starlights for a number of stars for which these two components have 
been estimated (\citet{uv3} and the Appendix of this paper).
The ratio of scattered to direct starlights increases with column 
density, as expected.
The effect of gas extinction (the exponential of $1/\lambda^{4}$ in 
Table~\ref{tbl:fit}) on the fits of the reduced spectra 
is small under $1/\lambda \sim 6\,\mu\rm m ^{-1}$, meaning that for 
this sample of stars $P_{cs}/F_{0}е$ is in the range $[0.1,\,100]$ 
around $1500\,\mu\rm m$.
For $N_{H}=10^{21}\rm cm^{-2}$ ($E_{uv}\sim 0.2$) the observed $P_{cs}/F_{0}е$ 
is between 0.1 and 10 (Figure~\ref{fig:rap}), thus substantially smaller than in the 
idealized conditions of sections~\ref{cs} and \ref{an}.
$P_{cs}/F_{0}е$ can grow with larger values of $E(B-V)$, but it 
cannot grow too much because the gas extinction term $e^{-g/\lambda^{4}}$
of equation~\ref{eq:fit1} will then become important.
\section{Energy conservation} \label{enercons}
As long as Fresnel's theory is correct, there is no doubt that 
$N_{H}S_{\lambda}$ atoms locked in the first Fresnel zone will give 
a scattered flux, at the position of the observer,
given by equation~\ref{eq:mag}, and estimated in section~\ref{an}, 
very much greater than the unreddened flux of the source.
Because of energy conservation, we must have: 
$P_{S_{0}е}/(r_{d}^{2}P_{cs})\gg 1$.
$P_{S_{0}е}$, for a cloud half way between the source and the 
observer, and 100~pc away from the sun, can be estimated by:
$4\sigma N_{H}S_{\lambda}F_{0}\sim 4\,10^{11}F_{0}$ 
(equation~\ref{eq:ps0}), if 
$E(B-V)=0.2$ and for $\lambda= 1500\,\rm\AA$.
From Figure~\ref{fig:rap}, $P_{cs}$ is between $F_{0}$ andе $10F_{0}$ at $\lambda= 
1500\,\rm\AA$ ($1/\lambda=6\,\mu\rm m^{-1}$), for $E(B-V)<0.3$.
The condition imposed by energy conservation is then: $r_{d}\ll 
2\,10^{5}\,\rm cm$.

The power per unit surface scattered by $N_{H}S_{\lambda}$ atoms in the 
first Fresnel zone is sufficient to explain the deviation from the linear extinction by 
large grains observed in the UV.
However, interstellar clouds are certainly not limited to a small space 
between the star and the observer, and for an homogenous medium
a unit surface at the position of the observer 
should not receive more power than $10^{-4}\,еF_{0}$ 
(for $N_{H}\sim 10^{21}\,\rm cm^{-2}е$), and in no way more 
than $F_{0}$е (section~\ref{us}).е
The implication of this apparent contradiction is that interstellar 
clouds are not homogenous at scales larger than $r_{0}$, which means 
that density fluctuations must exist at scales smaller than a few 
hundred kilometers.

Reality is then in between these two extreme approximations, of a perfectly 
homogenous medium on the one hand, of a concentration of gas in the 
first Fresnel zone (and nothing around) on the other.
This is further justified by the observational orders of magnitude found in 
section~\ref{obs} which are significantly smaller than the 
theoretical predictions of section~\ref{an}, and much higher than what the 
brightness of an homogenous medium can be presumed to be.
\section{Wavelength dependence of the scattered light} \label{wave}
The maximum flux observable from coherent scattering from 
the first Fresnel zone at wavelength $\lambda$ varies as $1/\lambda^{2}$ 
(Equation~\ref{eq:pcm}).
However, coherent scattering occurs from well defined interstellar structures 
which each will give 
a $1/\lambda^{4}$ dependence (Equation~\ref{eq:pc}), 
as it is indicated by observation \citep{uv3,uv6}.е

It is also worth noting that the depth of the cloud along the line 
of sight (which was not considered in the preceding sections) will not change the 
$1/\lambda^{4}$ wavelength dependence.
As far as $d_{0}$ (and $l_{0}$)е does not vary appreciably across the cloud, the 
waves scattered by atoms close to the line of sight, regardless of 
their exact projected position on the star-observer axis, 
reach the observer with identical amplitudes and phases, and add positively.
If $d_{0}$ varies appreciably across the cloud the phases at the 
position of the observer will remain the same, though a sum over different amplitudes will 
have to be considered; but
the $1/\lambda^{4}$ dependence of the scattered light will be unchanged.
\section{Dependence of the scattered light on geometry} \label{geom}
The scattered coherent light from the cloud depends on geometry 
through the dependence of $S_{0}$ on distances, and through the small 
scale organisation of interstellar matter.
For instance,
if the scattering medium is close to the star, 
path-lengths' differences will be large even very close to 
the direction of the star; coherent scattering is restricted to a 
quasi-null surface around the complete forward direction 
($S_{0},\,\rm as\, S_{\lambda}е\rightarrow 0$ if $\d_{0}\rightarrow 0$, 
equation~\ref{eq:sl}), and $P_{cs}$ should tend to 0.
\section{Stars with circumstellar dust} \label{geo}
We expect (section~\ref{geom}) stars with circumstellar dust
to have less scattered starlight contamination than if 
they were farther behind a cloud.
If, as I believe \citep{uv8}, the bump is associated with the scattered 
light component, it follows that stars surrounded by interstellar 
matter should have a smaller bump than expected from their reddening.
This is exactly the conclusion \citet{sitko} reach from a sample of hot stars 
with a dust shell.

In the next sections I discuss the case of planetary nebulae 
(P.N.), and the shell stars of \citet{sitko} article. 
Two methods can be used to detect an absence of scattered light in the 
UV spectrum of these object.
One way is to prove [as it was done in \citet{uv5} for directions of 
very low reddening] that the extinction law remains the same in the 
visible and in the UV:
this is the case if the UV reduced spectrum of the object 
exponentially decreases as 
$e^{-2E_{uv}е/\lambda}$ ($\lambda$ in $\mu$m, see \citet{uv2}), with 
$E_{uv}\sim E(B-V)$.
Another way is to check the presence of a $2200\,\rm\AA$ bump:
the study given in \citet{uv5}, on the extinction in very light 
interstellar media, shows that the threshold of $E(B-V)$ 
above which a bump generally 
appears should be around 0.05~mag.; 
it is also above this value that the far-UV part of the extinction curve 
will in general no more be fitted by a straight line.
The absence of a bump for a reddening much larger than 0.05~mag. indicates 
that the scattered light component is negligible.
\begin{table*}[p]
\caption[]{IUE planetary nebulae with no apparent bump}		
       \[
    \begin{tabular}{|l|c|c|c|c|c|}
	\hline
\multicolumn{6}{|c|}{IUE planetary nebulae with no apparent bump}\\
\hline
name & $E(B-V)^{(1)}$ &$E(B-V)_{ci}е^{(2)}$  & 
$d_{ph}^{(3)}$ & $d_{tr}^{(4)}$ & IUE file $^{(5)}$\\ 
\hline
NGC6905, HD193949& 0.651  &  0.166 &  5.4  &- &LWR10067\\
NGC5189, HD117622  &  0.6 & b=-3.45  &  1.5  &- & LWR07171\\
BD$-22^{\circ} 3467$, A35&  0.5a & 0.112  &    &0.134 &LWP22520\\
NGC6818, HD186282 & 0.217  & 0.162  &  2.5  & - & LWR01608\\
NGC6720, HD175353&  0.2 & 0.095  &  400  & - & LWP20393 \\  
NGC3211, HD89516 & 0.20  &  b=-4.87 &  3.4  & - & LWR03203\\
FB138  & 0.2a  & 0.074  &    &0.243 &LWR12596\\
NGC2392, HD59088 & 0.15 (0.17a)  &  0.051 &  3.5  & 1.1 & LWP24370\\
NGC650, HD10346,& 0.14  & 0.25  &  0.5  & - & LWP20915\\
NGC6853, M27  &  0.126 &  b=-3.70 &  0.4 & -  & LWR05515 \\
NGC7662, HD220733  & 0.105  &  0.139 &  1.6  & - & LWP20683 \\
NGC2371  &  0.1 & 0.046  &  5.3  & - & LWR04210  \\
IC2448, HD78991 & 0.08  &  0.101 & 4.4 &- & LWR02756\\
IC4593, A27  &  0.04 &  0.058 &  3.9  & - & LWR05705\\
NGC4361, HD107969  &  0.03 &  0.041 &  1.6  & - & LWR15879\\
NGC6210, HD151121 &  0.03 & 0.054  &  2.5  & - & LWR09422\\
NGC246 & 0.02a  & 0.027  &    &0.63H  & LWR06806\\
NGC6826, HD186924&  0.02 & 0.11  &  1.2  & - & LWR12580\\
NGC7293, Helix  &  0.02 &  0.292 &  0.2  & - & LWR08942\\
\hline
\end{tabular}    
    \]
\begin{list}{}{}
\item[$(1)$] Reddening of the P.N..
$E(B-V)\sim 0.7c$ is calculated from the `$c$' (the logarithmic 
extinction in H$\beta$) column of Table~1 in \citet{cazetta00};
$a$ after the magnitude, indicates that $E(B-V)$ comes from 
\citet{acker98}.
\item[$(2)$] Large scale value of $E(B-V)$ from \citet{schlegel} 
(http://nedwww.ipac.caltech.edu/forms/calculator.html).
Gives the reddening due to interstellar cirrus in the direction of 
the P.N., for $|b|> 5$.
When $|b|< 5$, the galactic latitude replaces $E(B-V)_{ci}е$.
If the P.N. is behind the cirrus, $E(B-V)_{ci}$ is the reddening 
due to foreground dust, excluding the material surrounding the star.
\item[$(3)$] Photometric distance (kpc) from \citet{cazetta00}.
\item[$(4)$] Trigonometric distance (kpc), from Hipparcos, when 
available. `$\infty$' 
is for negative Hipparcos parallaxes.
\item[$(5)$] IUE number of one IUE long-wavelength spectrum that can be 
used to prove the absence or the presence of a $2200\,\rm\AA$ bump.
\end{list}
\label{tbl:pn0}
\end{table*}
\clearpage
\begin{table*}[p]
\caption[]{IUE planetary nebulae with a bump. See Table~\ref{tbl:pn0} 
for the meaning of each column.}		
       \[
    \begin{tabular}{|l|c|c|c|c|c|}
\hline
\multicolumn{6}{|c|}{IUE planetary nebulae with a bump}\\
\hline
name & $E(B-V) $ &$E(B-V)_{ci}$  & $d_{ph}$ & $d_{tr}$&IUE file\\ 
\hline
Hen2-99  &  0.784c & 0.50  &  4.2  &-&LWP15068 \\
NGC6751, HD177656  &  0.76 &  0.517 &  3.9  & - &LWR10775\\
IC4997, HD193538 & 0.75 & 0.139  &  1.2  & -&LWR11011\\
NGC6629, HD169460  & 0.63  & 0.606  &  2.1  &- &LWP15329\\
NGC6567, HD166935  & 0.52  &  b=-0.65 &  1  & - &LWP23708\\
NGC40,  HD826  & 0.5 (0.3a)  & 0.39 &  1.5  & $\infty$&LWR15104 \\
HD167362, SwSt1  &  0.50a & 0.274  &    &0.112 &LWR04804\\
NGC1514, HD281679  & 0.45a  & 0.679  & - & 0.185 &LWR01279\\
Hen2-108  &  0.385c & 0.44  &  6.4  & - &LWR10776\\
NGC5315, HD120800  &  0.42 & b=-4.40  &  4.3  &- &LWR06963\\
IC4634, HD153655  & 0.39  & 0.379  &  5.7  &- &LWP13828\\
IC2501, HD83832  & 0.37  &  0.464 &   7.1 & -&LWR12566\\
IC2165, HD44519  & 0.36  & 0.406  &  2.6  &-&LWR10507\\
BD+303639, HD184738  & 0.32 & 0.403  &  2.3  &$\infty$ &LWR07333\\
NGC2867, HD81119  & 0.30  & 0.4  & 3.3   & -&LWR04510\\
NGC5882, HD135456  & 0.27  & 0.297  &  2.5  &- &LWR12599\\
NGC6572, HD166802  & 0.259  &  0.252 &  2.7  & - &LWR07421\\
Hen2-438, HD184738& 0.24a &0.403   &      &$\infty$ &LWR07333\\
IC418, HD35914  &  0.22 & 0.221  &  1.9  & - &LWR02741\\
NGC6891, HD192563  &  0.21 &  0.185 &  3.5  & -&LWP23363\\
IC2149,HD39659  &  0.18 & 0.234  &   2.1 & - &LWP14488\\
NGC6153, HD148687  &  0.13 & 1.383  &  4.4  &- &LWP13841\\
IC3568, HD109540  &0.13   & 0.137  &  3.4  &- &LWR10509\\
\hline
\end{tabular}    
    \]
\label{tbl:pn1}
\end{table*}
\clearpage
\section{Stars with circumstellar dust; planetary nebulae} \label{pn}
For this section I will use the large dataset of planetary nebulae 
(P.N.) from 
Table~1 in \citet{cazetta00}, and the P.N., observed by Hipparcos, of 
\citet{acker98}.
This sample of P.N. covers a large part of class~70-71 IUE objects 
(planetary nebulae observed with and without the central star).

In Table~\ref{tbl:pn0} and Table~\ref{tbl:pn1}, these P.N. have been 
classified according to whether they have a bump at $2200\,\rm\AA$ or 
not.
P.N. are presented in decreasing order of reddening.

The striking difference of P.N. with no bump and P.N. with a bump, is 
that the reddening of the former is on the average much lower than 
that of the latter.
This increase of reddening of P.N. with a bump compared to that of P.N. 
with no bump correlates with an increase of the `large scale' 
reddening, $E_{ci}$, due to interstellar cirrus on the line of sight.
Consequently, the difference of reddening between the two classes of P.N. 
(from Table~\ref{tbl:pn0} and Table~\ref{tbl:pn1}) will be attributed to 
a difference of foreground extinction.
If this part of the reddening could be identified and substracted, it is 
probable that Table~\ref{tbl:pn1} and Table~\ref{tbl:pn0} would be alike.

Thus, if it is assumed that the two sets of P.N., P.N. with a bump 
and foreground dust from Table~\ref{tbl:pn1}, 
and P.N. without an apparent bump and negligible foreground dust from 
Table~\ref{tbl:pn0}, differ because of the foreground extinction but
have the same intrinsic properties on the average, it will be concluded that:
\begin{itemize}
    \item  The reddening of a P.N. due to circumstellar dust is generally low. 
    For most P.N., intrinsic $E(B-V)$ should not exceed 0.1-0.2~mag.. 
    Except for a few cases, larger extinctions 
    are due to foreground dust.
    \item  A $2200\,\rm\AA$ bump in a P.N. is due mainly to foreground dust, 
    not to the surrounding material.
\end{itemize}е
That the intrinsic $E(B-V)$ should be low, 0.1-0.2~mag 
at most, reflects a general feeling, already expressed in the 
literature \citep{phillips84}.
According to \citet{phillips98}, exceptions are compact nebulae of small radius ($< 0.1$~pc).
This may be the case of A35, NGC6905 and NGC5189 in 
Table~\ref{tbl:pn0}, and of IC4997 and HD167362 in Table~\ref{tbl:pn1}, for 
which the intrinsic reddening is apparently large, since the large 
scale reddening is low in these directions (though local enhancements of 
the foreground column density can not be excluded).
A35, NGC6905  are cited by 
\citet{phillips84} (see his Figure~3) as two among the few P.N. with 
higher intrinsic reddening than the average.

The intrinsic reddening of A35, NGC6905 and NGC5189 should be larger 
than 0.5~mag.
The absence of an apparent bump in these nebulae, as 
for other P.N. with intrinsic $E(B-V)$ larger than 0.05~mag, 
is, according to the framework developped in this paper, due to an absence of 
scattered light contamination, because of the proximity of the star 
and of the obscuring material (section~\ref{geom}).

To conclude this section on the reddening of planetary nebulae, it is worth mentionning 
the article of \citet{stasinska92} (and references therein), which compare the interstellar 
extinction of P.N. by means of photometry and a standard extinction 
law, and by means of radio measurements.
There seems to be a systematic difference (always in the same way) 
when the reddenings found by the two methods are compared.
This difference increases with reddening.
According to  \citet{stasinska92}, it can only be due to a 
modification of the interstellar extinction law for clouds at larger 
distances than 2.2~kpc from the sun. 
At these distances, 
$R_{V}=A_{V}/E(B-V)$ would be smaller than in the vincinity of the sun.

Strictly speaking the results of  \citet{stasinska92} state that:
\begin{itemize}
    \item  The observed extinction law depends on the distances 
    ($l_{0}$ and $d_{0}$)ее from 
    the observer.
    \item  Standard extinction laws do not properly describe the interstellar 
reddening. 
\end{itemize}е
Both items agree with the ideas discussed so far.
That interstellar extinction varies with distance from the observer 
can hardly be explained by a modification of its 
properties, which would lead to a questioning case of geocentricism:
why should the properties of interstellar grains 
vary with distance from the sun?
\begin{table*}[p]
\caption[]{\citet{sitko} shell stars}		
       \[
    \begin{tabular}{|l|c|c|c|c|c|c|c|}
\hline
star& S.T.$^{(1)}$& $B-V\,^{(2)}$ &$E^{(3)}$ &$E_{ci}е^{(4)}$ & 
$d_{H}^{(5)}$&$  (\frac{{100}}{{60}})$\tiny{$_\mathrm{ci}$}\normalsize$^{(6)}$&
$ \frac{100_{\star}}{100_\mathrm{\tiny{ci}}}^{(7)} $\\ 
\hline
BD$+40^{\circ}\,4124$&B2 &0.77 & 1.00 &b=+2.77 &108 &1.7&0.9\\
BD$+61^{\circ}154$&B8 &0.55 &0.66 &b=-0.95 &300 &5.0&0.8\\
V380 Ori&A0 & 0.48&0.49 &4.528 & 269&4.5&0.7\\
HD259431& B5 &0.28 &0.43 &b=+0.67 &290 &4.5&0.5\\
HD45677& B2&0.04 & 0.28&0.307 & 355&5.0&0.2\\
AB Aur&B9-A0 & 0.12&0.15 &0.928 & 144&4.0&0.3\\
HD50138&B8-B9 & +0.01&  0.10&b=-3.14 & 290&4.5&0.4\\
HD190073&A0 &0.09 0& 0.10&0.118&5000&4.0&0.7\\
HD163296&A1V & 0.07& 0.05&b=+1.49 &122&4.5&0.9\\
\hline
\end{tabular}    
    \]
\begin{list}{}{}
\item[$(1)$] Spectral type.
\item[$(2)$] $B-V$ from \citet{sitko}; updated with Simbad and the 
Lausanne University database (http://obswww.unige.ch/gcpd).
\item[$(3)$] $E(B-V)$ deduced from spectral type and  \citet{fitzgerald70}.
The uncertainty on $E(B-V)$ should not exceed $\pm 0.05$~mag..
\item[$(4)$] Large scale value of $E(B-V)$ from \citet{schlegel}. 
For stars in the galactic plane, galactic latitude is given.
\item[$(5)$] Estimated distance (pc) from the sun (Hipparcos).
\item[$(6)$] Infrared $\frac{100\,\rm\mu m}{60\,\rm\mu m}$ color, 
around the star, of the interstellar cirrus. This is the slope of the linear regression 
between the $100\,\rm\mu m$ and the $60\,\rm\mu m$ images, on a 
$2^{\circ}\times 2^{\circ}$ field centered on the star.
\item[$(7)$] Ratio of the $100\,\rm\mu m$ emission at the position of 
the star, to what should be the $100\,\rm\mu m$ emission of the 
cirrus, estimated from the $60\,\rm\mu m$ emission at 
the position of the star, and the linear regression found between the 
$100\,\rm\mu m$ and $60\,\rm\mu m$ emission (item (6)).
If the reddening of a star is mainly due to circumstellar dust, 
$\frac{100_{\star}}{100_\mathrm{{ci}}}$ should be small.
On the contrary, if the extinction is due to foreground dust 
$\frac{100_{\star}}{100_\mathrm{{ci}}}$ will be close to 1. 
\end{list}
\label{tbl:sitko}
\end{table*}
\clearpage
\section{Stars with circumstellar dust; \citet{sitko} shell stars} \label{sitko}
The first columns of Table~\ref{tbl:sitko} list the main 
characteristics of  the shell stars used in \citet{sitko}.
In this paper, the authors show that when intrinsic and foreground 
reddenings can be separated, the bump feature associated with the intrinsic 
reddening is less than expected for a `normal' star extinguished by the 
same amount of foreground interstellar matter.

Stars  in Table~\ref{tbl:sitko} are ordered by decreasing $E(B-V)$.
HD44179, the Red Rectangle, was excluded from the table, since we know 
today that we do not observe the star, hidden behind a thick torus of 
dust, but scattered starlight escaping from the edges of the torus 
\citep{waelkens}.
ZCMa (HD53179) and HD31648 were also discarded 
because the noise level of their LWP and/or SWP 
spectra is nearly as high as the signal itself.

All stars in Table~\ref{tbl:sitko} have cirrus on their line of sight. 
Column $E_{ci}$ is the average opacity of the cirrus in the 
direction of each star \citep{schlegel}  and column $I_{100}/I_{60}$ 
its mean infrared color.
Since the extinction due to these cirrus can be important, a major 
problem is to separate the reddening by circumstellar dust from the 
one due to the cirrus.
\subsection{Separation of intrinsic and foreground reddenings} 
\label{sitkored}
From the mean linear relation which exists between the $100\,\rm \mu m$ and 
the $60\,\rm \mu m$ emissions (seventh column of Table~\ref{tbl:sitko}) 
in the vicinity of each star, we can calculate what should be the $100\,\mu$m 
emission of the cirrus at the position of the star, $I_{100,ci}$, for 
the $60\,\mu$m surface brightness measured by IRAS at the same position.
The comparison of $I_{100,ci}$ to  $I_{100,\star}$,
the measured $100\,\mu$m surface brightness at the position of the 
star, gives an indication on the respective proportions of foreground 
and intrinsic reddenings:
if the proportion of foreground reddening is large 
$I_{100,\star}/I_{100,ci}$ should be close to 1, while a large proportion 
of intrinsic reddening should imply a high $60\,\mu$m surface brightness 
at the position of the star (because of the heating of the dust surrounding the star); 
thus a low $I_{100,\star}/I_{100,ci}$.
 
Additional indications on the origin of the reddening of the stars 
will come from the distance; from the comparison of $E_{ci}$ with the 
reddening of the stars; and from the level of this reddening, since 
intrinsic reddenings, as for P.N., can be presumed not to exceed 0.1-0.2~mag. 

These criteria applied to Table~\ref{tbl:sitko} permit the separation 
of BD$+40^{\circ}4124$, BD$+61^{\circ}4124$, V380~Ori, and HD190073, 
on the one hand, from HD45677, AB~Aur, and HD50138 on the other.
The cases of HD259431 and of HD163296, both at low galactic latitude, are more difficult to fix.
HD259431 has an intermediate $I_{100,\star}/I_{100,ci}$ value, and 
its reddening is larger than what can be expected for a normal 
intrinsic one, which may be due to a 
better equilibrium between intrinsic and foreground extinctions.
For HD163296 $I_{100,\star}/I_{100,ci}$ is close to 1 but this 
pre-main sequence star is close to the sun, and its reddening is low 
and may be due to the dust surrounding the star.

The main source of reddening for the former stars, BD$+40^{\circ}4124$, 
BD$+61^{\circ}4124$, V380~Ori, and HD190073, is probably 
foreground dust.
If so, the low foreground reddenings ($<<0.1$~mag.) \citet{sitko} find for 
V380~Ori and for HD190073 are underestimated.
This should also be the case for BD$+61^{\circ}4124$, for which 
\citet{sitko} give $E_{ci}\sim 0.3$~mag..
The proportion of inrinsic reddening for 
HD45677, AB~Aur, and HD50138, is larger.
\subsection{The bump feature and intrinsic reddening} 
\label{sitkobump}
\citet{sitko} tried to correct the extinction curves of
AB~Aur, HD50138, HD259431, and HD45677, for their foreground 
reddening, using a standard extinction law.
This correction being done [Figure~9 of \citet{sitko}] the authors find a bump feature
(attributed to intrinsic reddening) much lower than expected 
from  a `normal' reddening by foreground dust.

This conclusion is supported by a direct analysis of the reduced spectrum 
of the stars (Figures~\ref{fig:fit} and \ref{fig:sitkofit}, 
Table~\ref{tbl:fit}).
There are five stars in Table~\ref{tbl:fit} with reddening less than 0.2~mag.
The four stars whose reduced spectrum can be fitted without a scattered 
light component are all \citet{sitko} shell stars.
The bump-like depression in the spectrum of these stars,  
centered around $2400\,\rm\AA$, is due to the Fe lines.
HD50138 and AB~Aur in particular have reddenings well above the 0.05~mag 
threshold. 
Their reduced spectrum can be compared to that of 
HD23480 (Figure~\ref{fig:fit}), of lesser reddening, but with a clear  $2200\,\rm\AA$ 
bump feature.

For HD259431 and HD45677, which have higher reddening,
intrinsic and foreground extinctions are mixed and conclusions from 
the size of the bump are not so easy to reach.
Comparison of their spectrum to HD62542 or to HD149757 ($\zeta$~Oph) 
is meaningless since these two stars are surrounded by dust \citep{cardelli88}.
\section{Effect of scattered starlight contamination on distance 
estimates} \label{dis}
A component of scattered light in the spectrum of 
reddened stars, if it is not identified and separated from the direct 
starlight, 
will give an impression of less extinction by large grains than it 
really is the case. 
A star for which the extinction is underestimated appears farther, if 
its distance is  estimated by means of photometric methods, than 
it really is. 
Consequently, photometric distances to reddened stars will be 
overestimated when compared to trigonometric distances
if the presence of scattered light is large 
enough to modify the estimate of the reddening of the star.

This result can be retrieved through analytical considerations, since the presence of scattered light 
introduces a term which grows with $1/\lambda$:
scattered light in the spectrum of reddened stars will always modify 
extinction curves by decreasing the visible slope 
[$\propto E(B-V)$], and therefore lead to an underestimate of the reddening.

The difference between the reddening estimated from observation, 
$E(B-V)_{obs}$, and the true $E(B-V)$, deduced 
from the analysis of the spectrum of HD46223 \citep{uv3}, is $\sim 0.1$~mag..
For the stars with high reddening used in \citet{uv6} (Figure~4 of \citet{uv6}),
$E(B-V)-E(B-V)_{obs}$ can be up to $0.2$~mag.
The corresponding errors on the distance modulus 
[$DM=5\log d-5 \sim V-M_{V}е-3E(B-V)$, $V$ and $M_{V}$ are the 
observed and absolute magnitudes of a star in the $V$-band] are $+0.3$~mag 
and $+0.6$~mag, meaning that estimates of distance by means of UBV photometry 
can be out by up to $\Delta d/d = +12\%$.
This relative error can be enhanced if the reddening is 
estimated from wavelengths shorter than the $V$-band (H$\beta$ for 
example), since scattered light contamination increases with decreasing wavelength.

Hipparcos distances have been compared to photometric distances
for some star clusters (see \citet{vl97, vl99, robichon99, 
stello01} and an article of  
Floor van Leeuwen at Hipparcos website), and for some planetary nebulae 
\citep{acker98,pottasch98}.
There is agreement between the two methods for some objects, 
but there are cases for which the two methods 
lead to different results, with a systematic trend for Hipparcos  
distances to be  smaller, by up to $10\%$, than the classical 
photometric methods.

Since the same particularity exists for star clusters and for P.N., the 
cause is probably not linked to a specific class of objects (such as young 
clusters, as argued by \citet{vl97}; see \citet{stello01} who 
also rejects this hypothesis).
It is not likely either that the reduction procedure of Hipparcos data be the 
cause of the discrepency;
it would then affect all Hipparcos data, while there is a good agreement between Hipparcos 
and other ground-based parallaxes \citep{harris97}.

I wondered if the presence of scattered light in the 
spectrum of reddened objects could influence the 
determination of their reddening and distance, and contribute to the differences 
observed between Hipparcos and photometric distances.
In general, since extinction is much enhanced in the UV, 
UV extinction curves will be a powerful tool in determining the true 
extinction of a star for which the distance is to be determined by 
means of photometry.
\section{Towards a fit of the extinction curve } \label{fit}
\begin{figure*}[p]
\resizebox{\hsize}{!}{\includegraphics{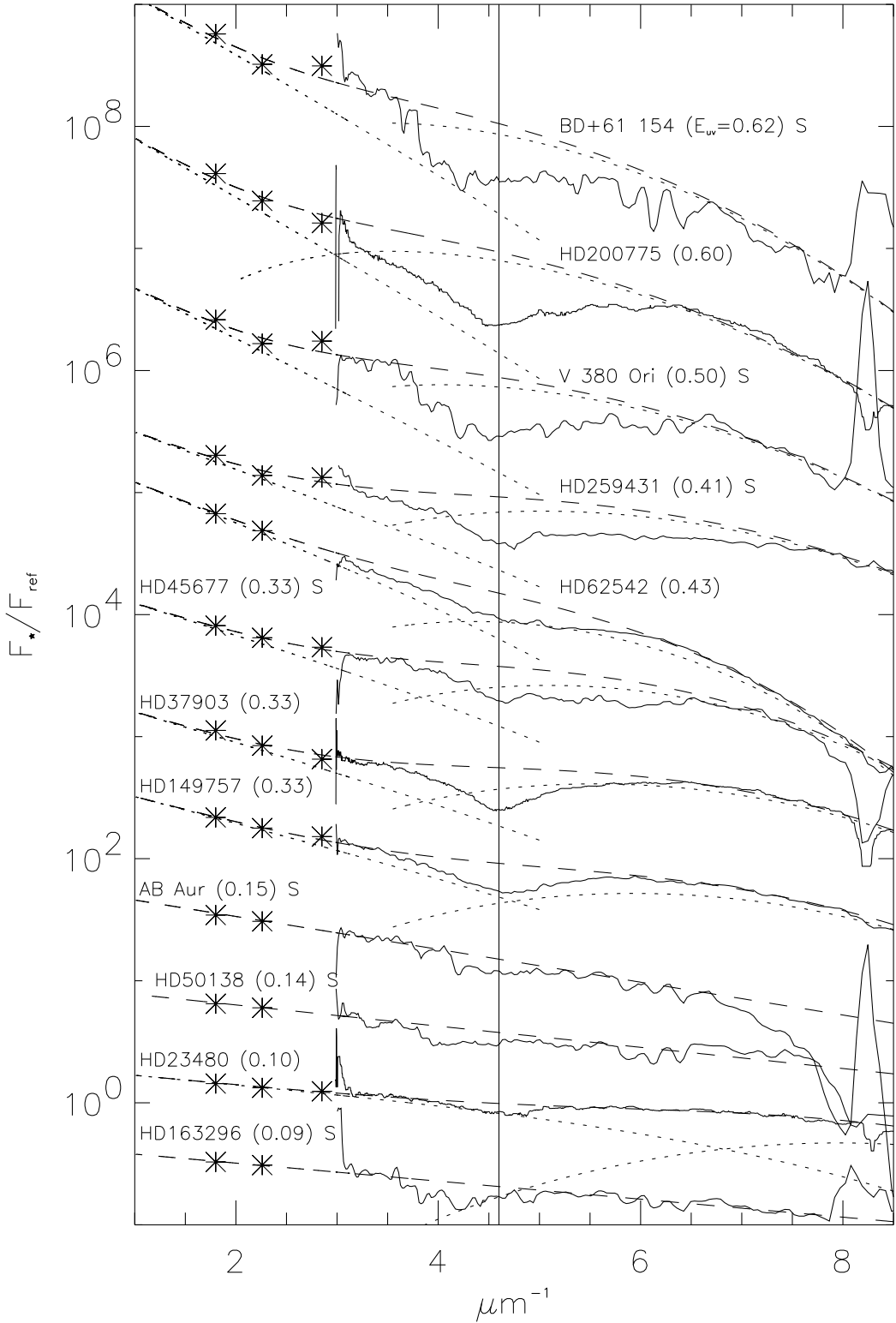}} 
\caption{Recapitulative plot of all the reduced spectra fitted in the 
Appendix.
Fits are represented as dashes and reduced spectra as solid lines. 
The spectra are scaled by an arbitrary factor.
$E_{uv}$ is the reddening found from the fit, which can be slightly 
different from $E(B-V)$ deduced from photometry (see 
Table~\ref{tbl:fit}).
Fits are given in Table~\ref{tbl:fit}. 
\citet{sitko} stars are labeled with an `S'.
}
\label{fig:fit}
\end{figure*}
\begin{figure*}[p]
\resizebox{\hsize}{!}{\includegraphics{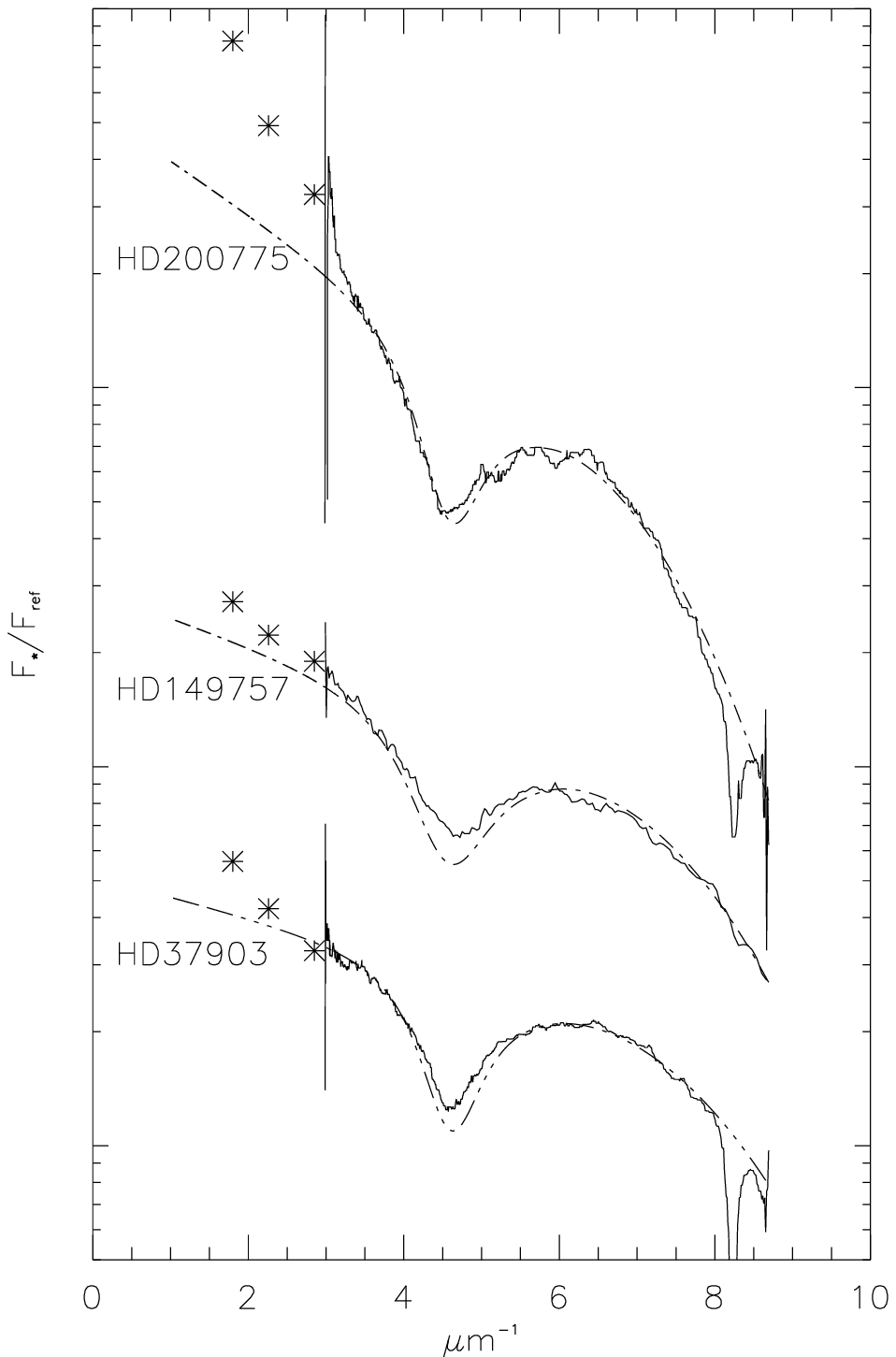}} 
\caption{\citet{fitz3} fits (dashes) for three reduced spectra of 
Figure~\ref{fig:uv2fit} (also represented on Figure~\ref{fig:fit}).
The fits are good for the UV part of the spectrum (less in the bump 
region) but diverge in the 
visible.
}
\label{fig:fitz}
\end{figure*}
\subsection{The standard theory framework } \label{stfit}
In the framework of the standard theory of extinction (i.e. if we do 
observe the direct light from reddened stars) the comparison of 
extinction curves, normalized to the same column density of large grains,
is necessary to understand the relationships between the different 
types of particles responsible for the overall extinction.
If the particles responsible for the UV extinction were the same large 
grains responsible for the visible extinction -or are in a constant proportion 
to these large grains- 
UV normalized extinction curves should superimpose well.
In the contrary extinction curves normalized  
in the visible will differ in the UV.

Although logical, this reasoning leads to a dead-end.
Extinction curves normalized by $E(B-V)$ have aleatory behaviors in 
the UV, 
meaning that the relations which clearly exist between the different parts of a 
non-normalized extinction curve [\citet{savage85}, or \citet{uv6} for 
the relation between visible and far-UV extinctions] are lost when 
extinction curves are normalized by $E(B-V)$;
the number of parameters necessary to fit normalized extinction curves grows to 6, 
in contradiction with the empirical finding of 
\citet{cardelli89}:
\emph{`The most important result presented here is that the entire mean 
extinction law, from the near-IR through the optical and 
IUE-accessible UV, can be well represented by a mean relationship 
which depends upon a single parameter\ldots
The deviations of the observations from the mean relation (e.g., 
Figs.~1 and 2) are impressively small.'} 
[\citet{cardelli89}, p.252-253. The `mean relation' of their Fig.~1
linearly relates, for UV wavelengths, $A_{\lambda}/A_{V}е$ and 
$R_{V}^{-1}=E(B-V)/A_{V}$, thus $A_{\lambda}$ to $E(B-V)$].

Moreover the \citet{fitz3} fits correctly reproduce the 
observations in the UV but diverge from it in the visible.
For three of the stars of Figure~\ref{fig:fit}, 
the fit coefficients were calculated by \citet{fitz2} or by \citet{aiello}.
Figure~\ref{fig:fitz} plots the reduced spectra of these stars along 
with their \citet{fitz3} fit.
The fits are good at UV wavelengths (except in the bump region) 
but clearly diverge from observation in the visible.
This absence of constraint of the \citet{fitz3} fit by the visible part 
of the extinction curve is a first explanation to the absence of relationship found between 
the  coefficients of the fit and $E(B-V)$ \citep{jenniskens93, aiello}.

In a recent paper \citep{fitz04}, E. Fitzpatrick
recognizes the limitations of the \citet{fitz3} fit,
and argues that the correlations found by 
\citet{cardelli89} between the different parts of non-normalized 
extinction curves is only due to a too small number of stars [29 
stars are used by \citet{cardelli89}].
This argument doesn't stand.
Firstly, \citet{cardelli89} stars are field stars, which were taken 
from \citet{fitz2} and had available near-IR photometry at 
the time the paper was written, with no 
particular assumption governing their choice.
Why shouldn't the following fifty or hundred stars follow the 
same relationships, except for possible 
deviations due to uncertainties on spectral types?
Secondly, even the 29 stars of the \citet{cardelli89} study present 
strong divergences in the UV if they are normalized by $E(B-V)$ (see 
Figure~6 of \citet{bless72}).
Lastly, the existence of a relationship, between the visible 
and far-UV parts of extinction curves, 
which is evident in the study of  \citet{cardelli89}, was independently 
retrieved by \citet{uv6}, on the basis of the physical ideas 
developed in the present paper. 

It is therefore the normalization process of extinction curves which is to be 
questioned, and not the number of stars used by \citet{cardelli89}.

If scattered starlight is added to direct starlight,
the light we receive from a reddened star will be composed of direct 
starlight, extinguished by large grains in proportion of the 
column density (thus proportional to $E(B-V)$), on the one hand, and of the 
starlight scattered by hydrogen, 
also extinguished by large grains, but proportional to the 
square of the column density [$E(B-V)^{2}е$], on the other.
Because of this difference in the dependence of the two components of 
the extinction curve on $E(B-V)$, normalization by $E(B-V)$ is meaningless. 
It can only increase the differences between extinction curves, and 
mask their relationships.
The normalization by $E(B-V)$ is the second reason, in addition to its limited wavelength 
domain of application, why no relation was ever found between the far-UV and 
visible parts of the \citet{fitz3} decomposition of the extinction 
curve. 
\subsection{A first step towards the fit of the extinction curve} \label{fitsca}
\begin{figure*}[]
\resizebox{\hsize}{!}{\includegraphics{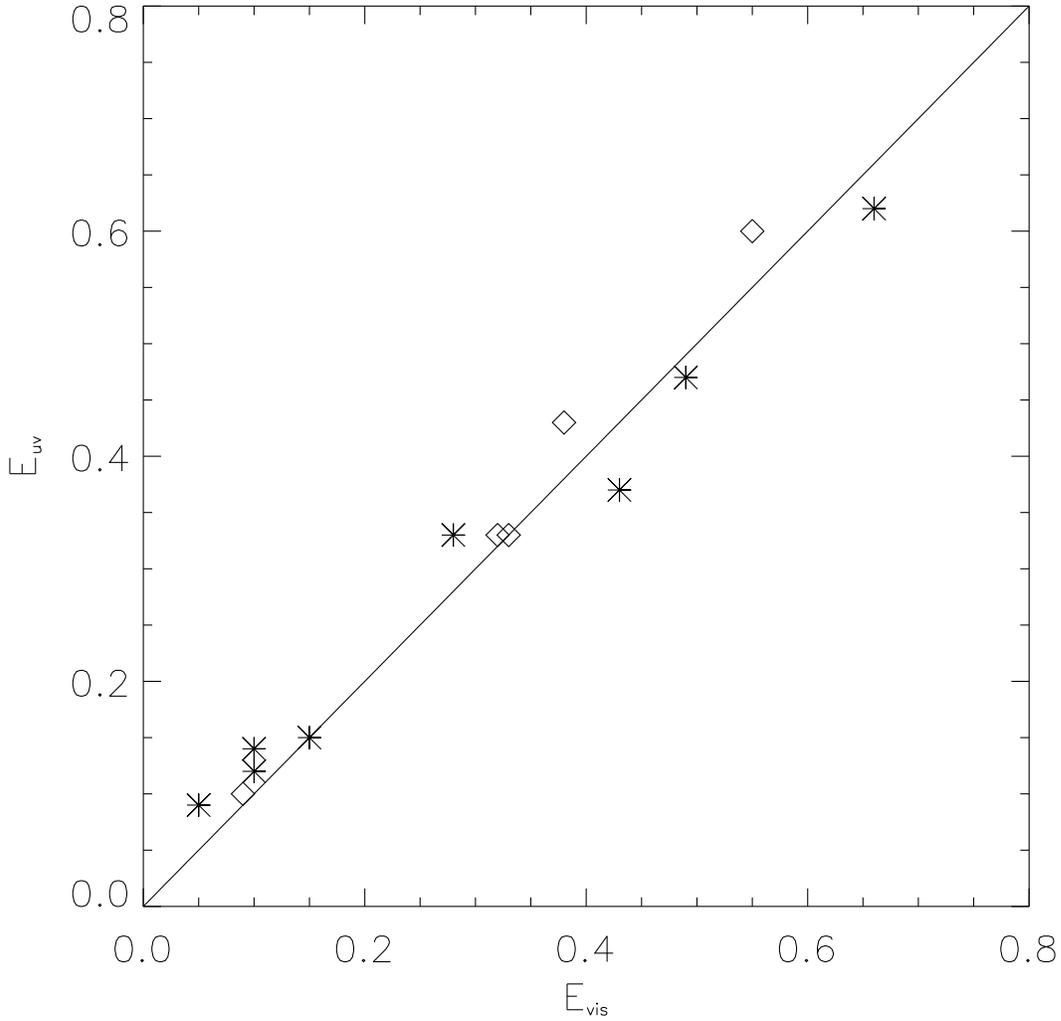}} 
\caption{Reddening estimated from the UV reduced spectra, $E_{uv}$,
versus the reddening estimated from the visible
photometry, $E_{vis}е$, for \citet{sitko} ($\diamond$) and \citet{uv2} 
stars ($*$).
}
\label{fig:euv}
\end{figure*}
\begin{figure*}[]
\resizebox{\hsize}{!}{\includegraphics{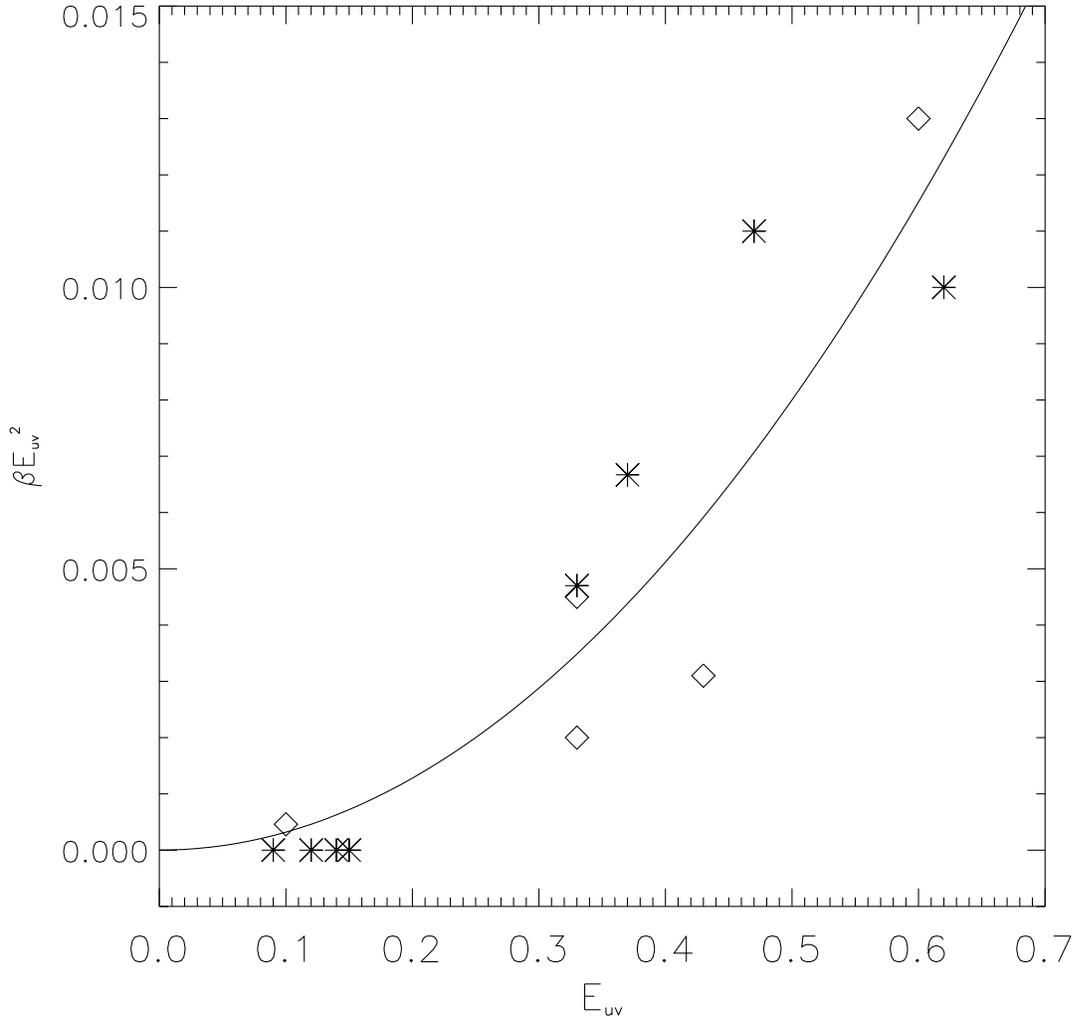}} 
\caption{$\beta E_{uv}^{2}$, the coefficient which defines the proportion of scattered 
light, versus $E_{uv}$ (equation~\ref{eq:fit1}) for \citet{sitko} ($*$) and 
\citet{uv2} ($\diamond$)
stars. $\beta E_{uv}^{2}$ roughly behaves as $E_{uv}^{2}$. 
}
\label{fig:beta}
\end{figure*}
\begin{figure*}[]
\resizebox{\hsize}{!}{\includegraphics{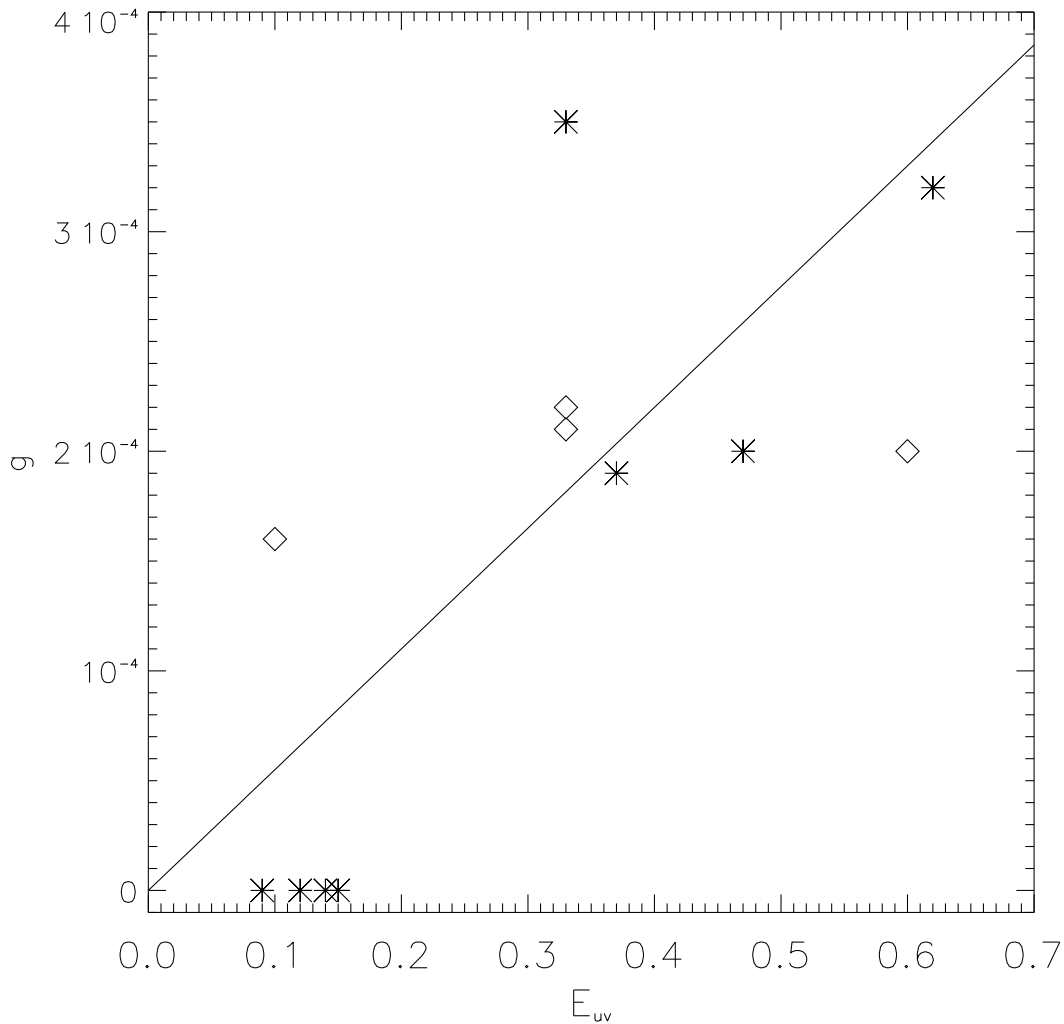}} 
\caption{$g$ (equation~\ref{eq:fit1}) 
versus $E_{uv}е$ for \citet{sitko} ($*$) and 
\citet{uv2} ($\diamond$)
stars. 
The fits are probably not precise enough to fix the exact dependence 
of $g$ on $E_{uv}$ [or $E(B-V)$].
The straight line gives $g\sim 0.00055 E_{uv}$.е
}
\label{fig:g}
\end{figure*}
\begin{figure*}[]
\resizebox{\hsize}{!}{\includegraphics{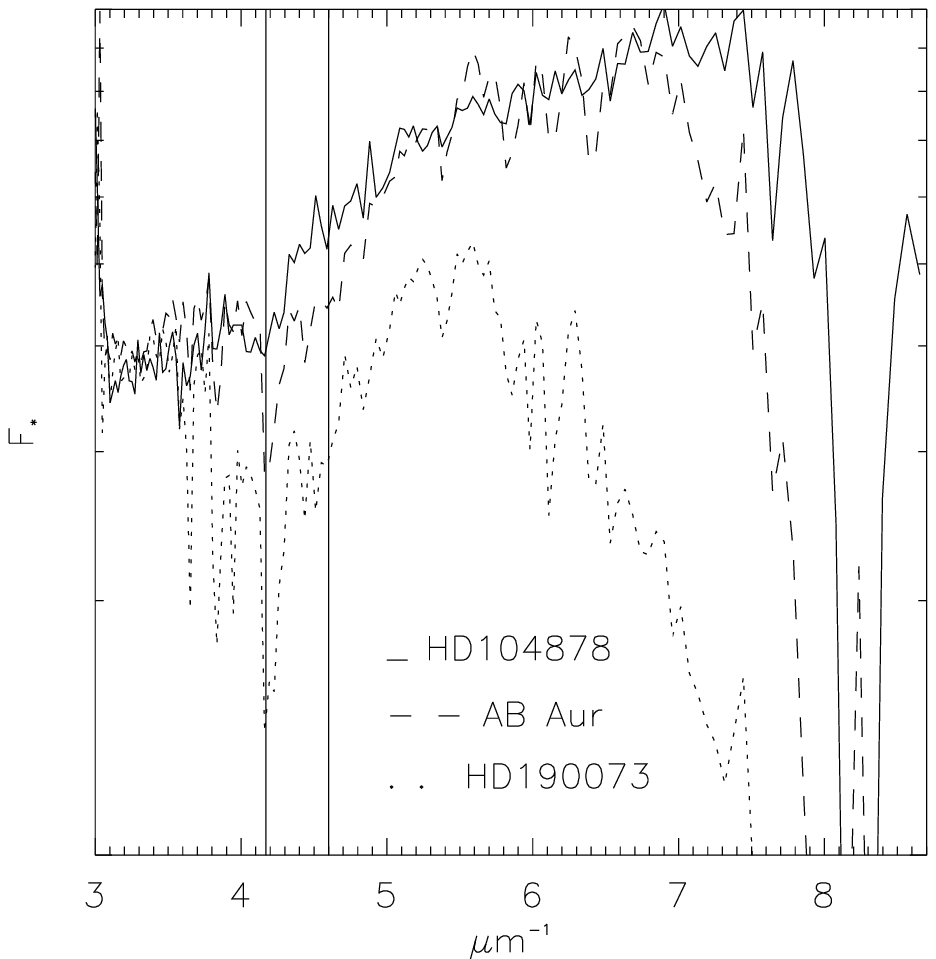}} 
\caption{The UV flux of the unreddened A0 star HD104878 is compared to the 
flux of AB~Aur multiplied by $e^{0.31/\lambda}$ (which accounts for 
the reddening of AB~Aur) and to the flux of 
HD190073  multiplied by $e^{0.25/\lambda}$.
The spectra superimpose well in the visible (not represented here, see 
Figure~\ref{fig:sitkofit} for AB~Aur) and in the 
near-UV.
In the far UV the spectra are difficult to compare because of the 
large Fe and Ly$\alpha$ absorptions. 
The positions of the Fe line and of the $2200\,\rm\AA$ bump are marked 
by vertical lines.
}
\label{fig:abaur}
\end{figure*}
Rayleigh scattering by hydrogen, with an additional extinction by 
large grains, should be fitted by a function 
proportional to \citep{rayleigh}:
\begin{equation}
  \frac{1}{\lambda^{4}}e^{-\frac{g}{\lambda^{4}}}e^{-2\frac{E_{s}}{\lambda}}
  \label{eq:gasca}
\end{equation}
The $e^{-2E_{s}/\lambda}$ term is for the extinction by large 
grains of the scattered light.
The $e^{-g/\lambda^{4}е}$ term accounts for extinction by the gas. 
$g$ should grow with the gas column density.
This term is usually considered as small compared with the extinction by 
large grains,  but it may gain in importance in the UV.
It must affect direct starlight as well as scattered light. 

Outside the bump region, the fit 
of the reduced spectrum of a reddened star should ressemble:
\begin{equation}
  e^{-2\frac{E_{d}е}{\lambda}}e^{-\frac{g_{d}}{\lambda^{4}е}}+
  E_{s}е^{2}еf(\frac{1}{\lambda})
  e^{-2\frac{E_{s}е}{\lambda}}e^{-\frac{g}{\lambda^{4}е}},
    \label{eq:fit0}
\end{equation}
$\lambda$ in $\mu$m.

Since direct and scattered light cross media which should be very much 
alike, we can suppose $g_{d}=g$ and $E_{d}е=E_{s}=E_{uv}$ ($E_{uv}$ is 
the true reddening, determined from the UV part of the reduced spectrum, 
which should be close to $E(B-V)_{obs}е$ deduced from 
photometry and spectral type).

$f(1/\lambda)$ contains the dependence of the scattered light, 
corrected for the extinction by large grains, on $1/\lambda$.
$f$ also depends on the distances to the cloud and to the star. 

Observations I have used \citep{uv3, uv6}, give an $f$ dependence 
on $\lambda$ as $1/\lambda^{4}$.
Therefore, the analytical expression of interstellar extinction 
outside the bump region should look like:
\begin{equation}
  e^{-2\frac{E_{uv}е}{\lambda}}e^{-\frac{g}{\lambda^{4}е}}\left(
  1+\beta \frac{E_{uv}е^{2}}{\lambda^{4}}е\right)
    \label{eq:fit1}
\end{equation}
$\beta$ depends on the distances to the cloud and to the star.

Figure~\ref{fig:fit} summarises all the fits calculated (from  
expression~\ref{eq:fit1}) and included in the appendix.
The spectra are ordered from top to bottom by 
decreasing reddening.
The expression of each fit is in Table~\ref{tbl:fit}.
Although there are uncertainties on these fits and they probably can be 
improved, the coefficients found already give an idea of the 
relationships which may exist between the parameters of expression~\ref{eq:fit1}.

Figure~\ref{fig:euv} confirms the relationship, already evidenced in 
\citet{uv6}, between the visible and the far-UV parts of the extinction 
curve.
$E_{uv}$, the value of the reddening in the 
direction of each star obtained from the analysis of the extinction 
curve, is plotted against $E_{vis}е$, the reddening determined from the star 
photometry. 
There is good agreement between the two values.

The $\beta E_{uv}е^{2}$ coefficient which fixes the importance of the scattered light
is plotted against $E_{uv}е$ in Figure~\ref{fig:beta}.
It roughly behaves as $E_{uv}е^{2}$, as expected, with $\beta \sim 0.035$ 
(neglecting the dependence on the geometry).

The gas extinction term $g$ affects the far-UV part of the fits, 
permitting a steeper curve in the far-UV.
It is not clear whether the steepness of the far-UV extinction 
curve is due to this term or to  
the wing of the Ly$\alpha$ absorption line.
AB~Aur (Figure~\ref{fig:sitkofit}) and HD190073 for instance have steep far-UV reduced spectra.
Since the reddening in these two directions is low it can be surmised 
that gas extinction is negligible -even in the UV- so that the 
steepness of the far-UV extinction curve in these directions is due 
to the wing of the Ly$\alpha$ absorption.
Comparison with the non-reddenned star HD104878 shows this is certainly the 
case for AB~Aur, and probably also for HD190073 (Figure~\ref{fig:abaur}). 
The absence of a clear relation in Figure~\ref{fig:g} plot of $g$ versus 
$E_{uv}$ will also be related to this difficulty of a precise 
determination of $g$.
The pertinence of a gas extinction term can probably also be 
investigated from a theoretical point of view by the comparison
of the extinction cross-section of the gas to the values of $g$ (a few $ 
10^{-4}$, if $\lambda$ is in $\mu$m) given in Table~\ref{tbl:fit}.е
\section{Conclusion } \label{con}
The hypothesis of the standard theory of extinction, that 
the UV light we receive from a reddened star is mainly the direct light from 
the star, and its corollary, the three component interstellar grain model, 
are not borne out by observation.

The only alternative to the standard 
theory is that the spectrum of a reddened star is contaminated by 
light scattered at small angular distances from the star.
The far-UV part of the extinction curve is observed to be less than 
the continuation of the visible extinction by large grains because of 
the importance of the scattered light component in the UV.

There is no need for particles such as PAH to explain the far-UV light we 
receive from a reddened star, with the probable implication that 
these particles do not exist in the interstellar medium.
This is not so much surprising since one can wonder how particles 
we are not even able to synthesize on earth, 
can proliferate in the cold and low density environment of an average interstellar cloud.

In the first part of the paper I have discussed several aspects of 
coherent scattering in 
the near forward directions, which is the unique reason I have found 
to explain the importance of the scattered light component 
in the spectrum of reddened stars \citep{uv4}.
There is up to date no theoretical treatment of this phenomenom, 
only briefly mentioned in some textbooks as the \citet{bohren}.
The reason might be the specific conditions necessary to its 
observation, which require 
large distances between the source of light, the scattering medium, and 
the observer; and
a non-homogenous (turbulent) medium.

Elementary calculations show that coherent forward scattering 
by hydrogen easily accounts for the order of magnitude found for the scattered 
light component of the spectrum of reddened stars, and will give the 
observed $1/\lambda^{4}$ wavelength dependence.
This scattering occurs within angles of $10^{-8}$'' or less from the 
direction of the stars, which represents a surface of radius a few $ 100\,\rm 
km$ for a cloud 100~pc away.
It implies that interstellar clouds still have structure at scales 
of a few hundred km.

If the cloud is close to the star ($d_{0}=0$) variations of the 
pathlengths for the scattered light will be large (compared to UV 
wavelengths) and random,
even at very close angular distance from the star.
The effect of coherence will be annihilated.
This latter result was announced by \citet{sitko}, who 
showed that hot stars with circumstellar dust have a $2200\,\rm \AA$ 
bump smaller than expected from their reddening.
\citet{sitko} result was 
extended to planetary nebulae (section~\ref{pn}).
It is concluded that circumstellar matter participates little or 
nothing to the $2200\,\rm \AA$ bump.
The bump, when present, is mainly due to a foreground cloud.
The extinction curves for the \citet{sitko} stars with $0.1<E(B-V)<0.2$ 
are a straight line from the visible to the UV: we observe the extinction of the direct 
starlight by the large grains (with a continuous extinction law) only.

The diminution of the $2200\,\rm \AA$ bump when extinction comes from  
circumstellar material emphasizes the 
link between the coherent scattered light and the bump.
It confirms an opinion I have already expressed \citep{uv4, uv8} that the bump is not 
an absorption feature but an interruption of the scattered light, or, 
as I believe today, a compensation between (1) the extinction by the gas of the 
direct starlight, and (2) the scattered light.

Standard extinction laws tend to underestimate the true extinction by 
large grains.
If the distance of a star is estimated by means of photometry, 
the unreddened flux of the star will be underestimated, 
and the star will be believed to be farther off than it really is.
The mismatch can be larger than $10\,\%$, and should be observed when 
photometric distances of reddened stars are compared to parallaxes.
This may be related to on-going problems \citep{vl04}
on the comparison between Hipparcos results to distances
determined by photometric means.

Based upon the dependence on $\lambda$ of the far-UV extinction curve 
in a few directions found in \citet{uv3,uv6}, I have discussed
what should be the fit of the extinction curve (outside the bump 
region).
This early attempt to give a physical meaning to the fit of the 
extinction curve was applied to the stars used in this paper and in 
\citet{uv2}.
The relationship between far-UV and visible extinctions found 
by \citet{uv6} is well confirmed.
The coefficient which determines the importance of the scattered light 
component roughly behaves as $N_{H}^{2}$, as expected.
The fit used in section~\ref{fit} and in the appendix introduces a gas 
extinction term $\propto e^{-g/\lambda^{4}е}$ which accounts in part 
for the steepness of the extinction in the far-UV.
It is not clear whether this effect is real or due 
to the wings of the Ly$\alpha$ absorption.

In section~\ref{stfit} I have also discussed the reasons why no 
relation was ever found using the \citet{fitz3} fit of the extinction 
curve, and more generally why this fit is not appropriate -despite a
new attempt, announced in \citet{fitz04}, to model simultaneously the spectra 
of the reddened and standard stars- to a correct representation of the extinction curve.

A last implication of this work, foreseen, and refused, by \citet{fitz04}, is that the 
true interstellar extinction law, and the average extinction properties of interstellar 
grains, are the same in all directions.
\appendix
\section{\label{appfit}Appendix: Fit of extinction curves}
\begin{figure*}[p]
\resizebox{\hsize}{!}{\includegraphics{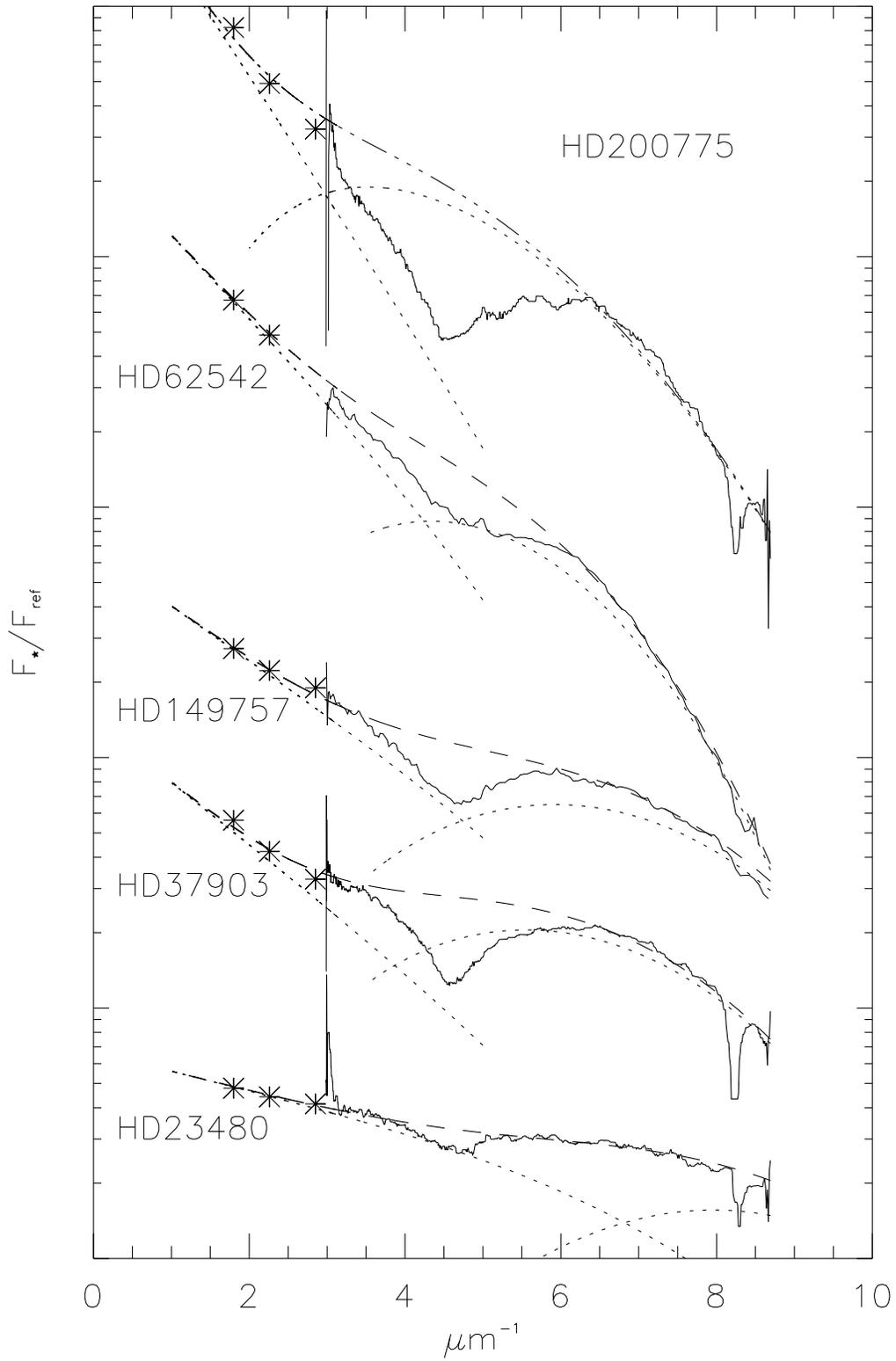}} 
\caption{Fit (dashes), outside the bump region, of the reduced spectra (solid lines) of \citet{uv2}, with an arbitrary 
logarithmic y-scaling.
The visible part of the spectra is represented by the phbotometric 
UBV points. Dotted lines are the direct and scattered starlight 
components.
The expression of each fit can be found in Table~\ref{tbl:fit} 
}
\label{fig:uv2fit}
\end{figure*}
The reduced spectra of the stars used in \citet{uv2} are fitted in 
Figure~\ref{fig:uv2fit}, using the general formulation of
section~\ref{fit}, equation~\ref{eq:fit1}.
There is not absolute uniqueness of each fit, and the relevance of the 
gas extinction parameter $g$ still remains to be proved 
(section~\ref{fit}).
In absence of more constraints I tried to determine a good 
fit to each of the spectra.
The fits extend from the visible to the far-UV, the visible spectra 
are replaced by the photometric UBV points retrieved from the 
Lausanne database (http://obswww.unige.ch/gcpd).

\begin{figure*}[p]
\resizebox{\hsize}{!}{\includegraphics{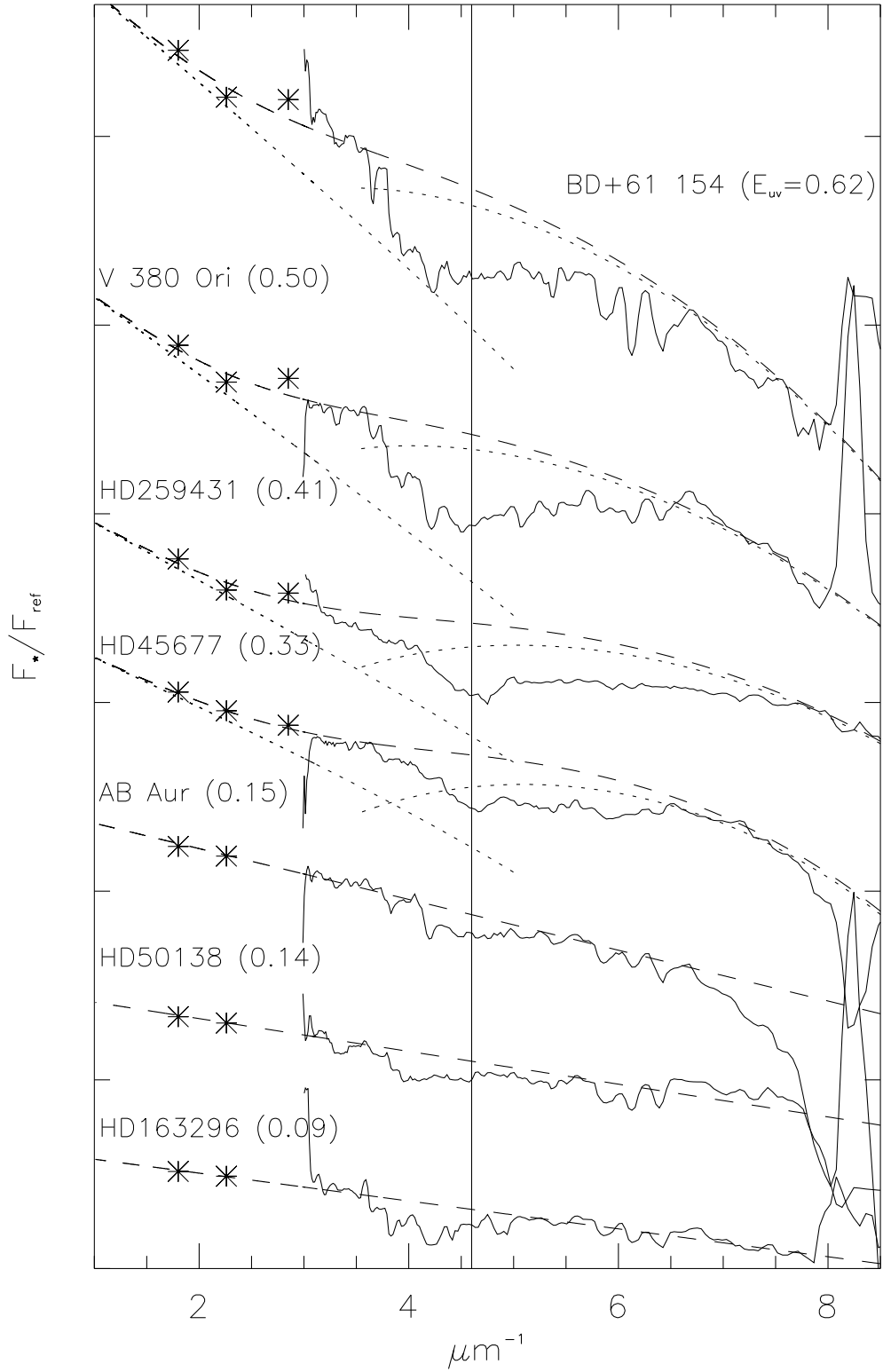}} 
\caption{Fits, outside the bump region, for the reduced spectra of \citet{sitko} stars.
Fit are represented as dashes and reduced spectra as solid lines.
y-axis is logarithmic.
The spectra are scaled by an arbitrary factor.
$E_{uv}$ is the reddening found from the fit, which can be slightly 
different from $E(B-V)$ deduced from photometry (see 
Table~\ref{tbl:fit}).
The analytical expression of each fit is in Table~\ref{tbl:fit}. 
}
\label{fig:sitkofit}
\end{figure*}

\begin{figure*}[p]
\resizebox{\hsize}{!}{\includegraphics{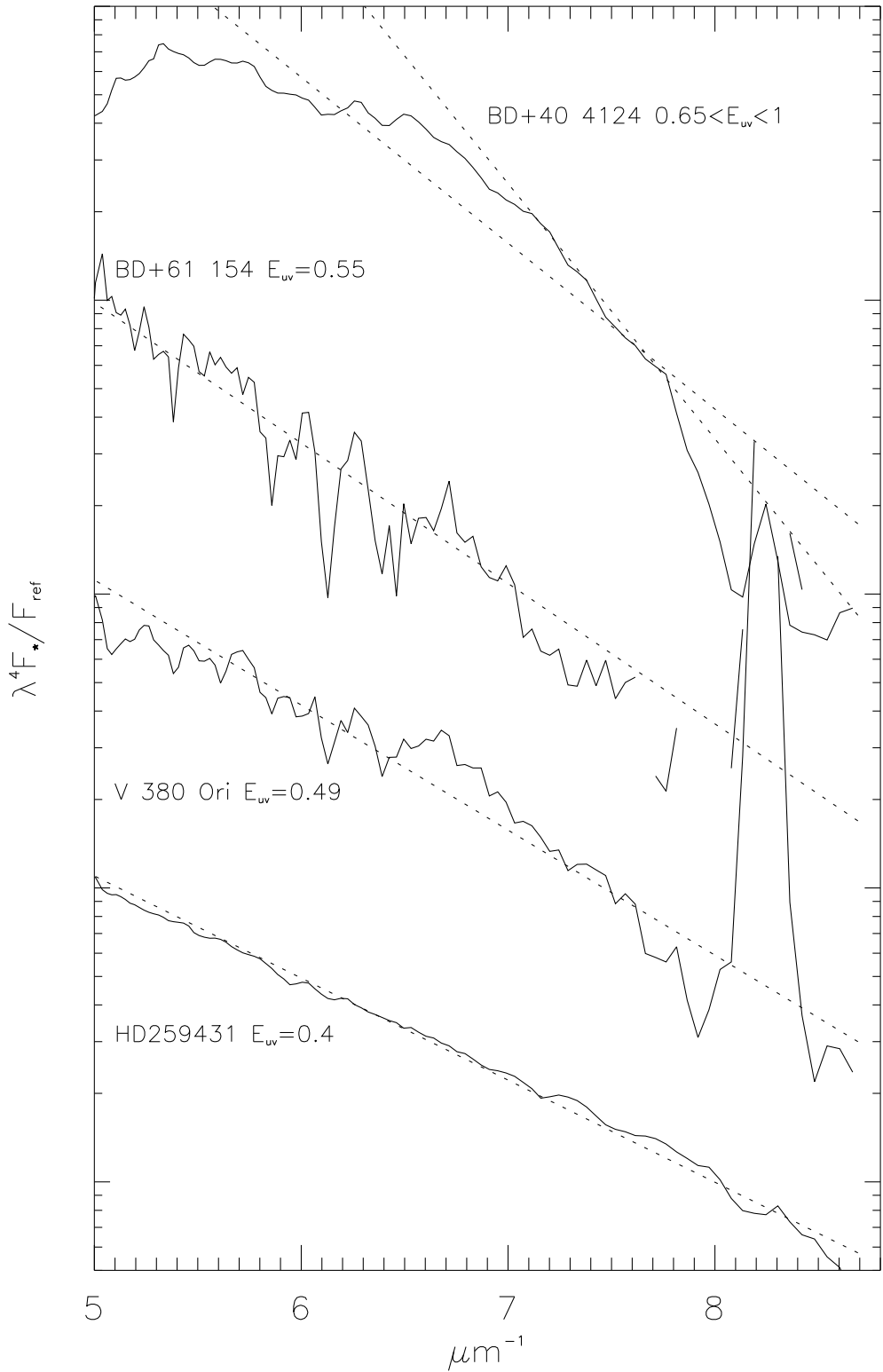}} 
\caption{Far-UV reduced spectra of \citet{sitko} stars of high reddening, 
multiplied by $\lambda^{4}$.
Each spectrum is well fitted by an exponential of $1/\lambda$, with an 
exponent $2E_{uv}$ close to $2E(B-V)$ (Figure~\ref{fig:euv}).
Only for BD$+40^{\circ}е\,4124$ is the slope difficult to determine 
because of the importance of the $2200\,\rm\AA$ bump and of the 
Ly$\alpha$ absorption.
} 
\label{fig:sitkogdtau}
\end{figure*}

The reduced spectra of \citet{sitko} stars can be classified, as I 
have done in previous papers, according to how far 
towards the visible the scattered component can be perceived.
HD163296, HD190073, HD50138, AB~Aur,
follow the same extinction 
law in the visible and in the UV (Figure~\ref{fig:sitkofit}), thus 
have a negligible scattered light contamination.
For HD45677, of moderate reddening, the visible extinction law 
extends down to the bump region, the scattered component appears as 
superimposed on the tail of the far-UV extinction by large grains.
The reduced spectrum of each of the four stars of higher reddening 
(HD259431, V380~Ori, BD$+61^{\circ}\,154$, BD$+40^{\circ}\,4124$)
multiplied by $\lambda^{4}$, is an exponential of $1/\lambda$ in the 
far-UV  (Figure~\ref{fig:sitkogdtau});
the exponent, $2E_{uv}$, is close to $2E(B-V)$ (Figure~\ref{fig:euv}).
The far-UV slope of BD$+40^{\circ}\,4124$, the star with the highest reddening, 
is difficult to determine, because of the small wavenumber interval 
between the bump on the short 
wavenumber side, and the Ly$\alpha$ line on the other side. 
Figure~\ref{fig:sitkofit} plots, as it was done for the stars of 
\citet{uv2}, the reduced spectra of \citet{sitko} stars (except for 
HD190073, see figure~\ref{fig:abaur}) and a possible 
fit deduced from equation~\ref{eq:fit1}.
\clearpage
\begin{table*}[h]
\caption[]{Fits used in Figures~\ref{fig:fit}, \ref{fig:uv2fit}, and \ref{fig:sitkofit}}		
       \[
    \begin{tabular}{|l|c|c|c|c|c|c|}
\hline
star& Ref$^{(1)}$& $E(B-V)^{(2)}$ &$E_{uv}е^{(3)}$ &$E_{ci}е^{(4)}$ 
&fit&source$^{(5)}$\\ 
\hline
BD$+61^{\circ}154$& HD23753 &0.66 & 0.62 &b=-0.95 & $\propto
e^{-\frac{1.25 }{\lambda}}e^{-\frac{0.00032 }{\lambda^{4}}е}
\left(1+ \frac{0.01}{\lambda^{4}} \right) $& S  \\
HD200775 & HD58050 & 0.55 & 0.60 & 11.946 &$ \propto
e^{-\frac{1.2}{\lambda}}e^{-\frac{0.0002}{\lambda^{4}}}
\left(1+\frac{0.013}{\lambda^{4}} \right)$  &uv2 \\
V380 Ori& HD104878 &0.49& 0.47 &4.528 &  $\propto
e^{-\frac{0.94}{\lambda}}e^{-\frac{ 0.0002}{\lambda^{4}}е}
\left(1+ \frac{0.011}{\lambda^{4}} \right)$ & S \\
HD62542 & HD32630 & 0.38 & 0.43&0.76 & $\propto
e^{-\frac{ 0.86}{\lambda}}e^{-\frac{ 0.0005}{\lambda^{4}}е}
\left(1+ \frac{0.0031}{\lambda^{4}} \right) $ & uv2 \\
HD259431& HD199081 &0.43 &0.37 &b=+0.67 &$\propto
e^{-\frac{ 0.74}{\lambda}}e^{-\frac{ 0.00019}{\lambda^{4}}е}
\left(1+ \frac{0.0067}{\lambda^{4}} \right) $ & S  \\
HD45677& HD31726  & 0.28 &0.33 &0.31 & $\propto
e^{-\frac{ 0.66}{\lambda}}e^{-\frac{0.00035 }{\lambda^{4}}е}
\left(1+ \frac{0.0047}{\lambda^{4}} \right)  $ & S\\
HD37903 & HD74273 &0.32 & 0.33 & 2.02 &$\propto
e^{-\frac{ 0.65}{\lambda}}e^{-\frac{ 0.00022}{\lambda^{4}}е}
\left(1+ \frac{0.0045}{\lambda^{4}} \right) $ & uv2 \\
HD149757 & HD214680 & 0.33 & 0.33 & 0.60 & $\propto
e^{-\frac{0.66 }{\lambda}}e^{-\frac{0.00021 }{\lambda^{4}}е}
\left(1+ \frac{0.0020}{\lambda^{4}} \right)$ & uv2  \\
AB Aur& HD104878 &0.15& 0.15 &0.928 & $\propto
e^{-\frac{ 0.31}{\lambda}}$ & S\\
HD50138& HD23753 &  0.10 & 0.14 &b=-3.14 & $\propto
e^{-\frac{ 0.28}{\lambda}} $  & S\\
HD190073$^{(6)}$& HD104878 & 0.10 & 0.12 &0.118& $\propto
e^{-\frac{ 0.25}{\lambda}}$  & S \\
HD23480 & HD215573 & 0.09 &0.10 & 0.64 & $\propto
e^{-\frac{ 0.20}{\lambda}} e^{-\frac{ 0.00016}{\lambda^{4}}е}
\left(1+ \frac{0.00046}{\lambda^{4}} \right) $ & uv2\\
HD163296& HD119765 & 0.05 & 0.09 &b=+1.49 &$ \propto
e^{-\frac{ 0.17}{\lambda}}$ & S  \\
\hline
\end{tabular}    
    \]
\begin{list}{}{}
\item[$(1)$] Standard star used to establish the reduced spectrum of 
the star. Each of these reference star was corrected for its slight reddening 
(see Table~\ref{tbl:ref}) according to \citet{uv5}).
\item[$(2)$] $E(B-V)$ from UBV photometry and spectral type. 
Corresponds to $E_{vis}$ in Figure~\ref{fig:euv}.
\item[$(3)$] $E_{uv}е$ deduced from the UV part of the reduced 
spectrum.
\item[$(4)$] Large scale value of $E(B-V)$ from \citet{schlegel}.
Galactic latitude for stars close to the galactic plane.
\item[$(5)$] `S' is for \citet{sitko} shell stars, uv2 for \citet{uv2}. 
\item[$(6)$] HD190073 was not represented on Figures~\ref{fig:sitkofit} 
and \ref{fig:fit}, but on Figure~\ref{fig:abaur}. 
\end{list}
\label{tbl:fit}
\end{table*}
\clearpage
\begin{table*}[]
\caption[]{Standard stars}		
       \[
    \begin{tabular}{|l|c|c|c|}
\hline
star & S.T.$^{(1)}$ & $B-V\,^{(2)}$ & $E^{(3)}$ \\ 
\hline
HD104878 & A0V &-0.011 &0.00 \\
HD119765& A1V &+0.007 & -0.01\\
HD199081& B5V &-0.136 & 0.02\\
HD214680 & O9V & -0.205& 0.10\\
HD215573& B6IV & -0.123 &0.02 \\
HD23753& B8V &-0.070 &0.04 \\
HD31726& B2V & -0.212& 0.03 \\
HD32630 & B3V & -0.146 &0.05 \\
HD58050 & B2Ve &-0.187 &0.05\\
HD74273 &B1.5V &-0.209&0.04\\
\hline
\end{tabular}    
    \]
\begin{list}{}{}
\item[$(1)$] Spectral type.
\item[$(2)$] $B-V$ from the 
Lausanne University database (http://obswww.unige.ch/gcpd).
\item[$(3)$] $E(B-V)$ deduced from spectral type and \citet{fitzgerald70}.
\end{list}
\label{tbl:ref}
\end{table*}
{}
\end{document}